\documentclass[prb,aps,amsmath,groupedaddress,twocolumn,eqsecnum]{revtex4}
\usepackage{amssymb}
\usepackage{amsmath}
\usepackage{graphicx}
\usepackage{amsbsy}
\usepackage{color}
\newcommand{\bq}{\begin{eqnarray}}
\newcommand{\eq}{\end{eqnarray}}
\newcommand{\bqn}{\begin{eqnarray*}}
\newcommand{\eqn}{\end{eqnarray*}}

\newcommand{\SW}{\text{SW}}

\begin{document}
%%%%%%%%%%%%%%%%%%%%%%%%%%%%%%%%%%%%%%%%%%%%%%%%%%%%%%%%%%%%%%%%%%%%%%%%%%%%%%%
%%%%%%%%%%%%%%%%%%%%%%%%%%%%%%%%%%%%%%%%%%%%%%%%%%%%%%%%%%%%%%%%%%%%%%%%%%%%%%%
%%%%%%%%%%%%%%%%%%%%%%%%%%%%%%%%%%%%%%%%%%%%%%%%%%%%%%%%%%%%%%%%%%%%%%%%%%%%%%%
\title{A Numerical Test of a High-Penetrability Approximation for the One-Dimensional Penetrable-Square-Well Model}

\author{Riccardo Fantoni}
\email{rfantoni@ts.infn.it}
\affiliation{National Institute for Theoretical Physics (NITheP), Stellenbosch 7600, South Africa}
\author{Achille Giacometti}
\email{achille@unive.it}
\affiliation{Dipartimento di Chimica Fisica, Universit\`a Ca' Foscari Venezia,
Calle Larga S. Marta DD2137, I-30123 Venezia, Italy}
\author{Alexandr Malijevsk\'y}
\email{a.malijevsky@imperial.ac.uk}
%\affiliation{E. H\'ala Laboratory of Thermodynamics, Academy of
%Science of the Czech Republic, Prague 6, Czech Republic and Institute
%of Theoretical Physics, Department of Chemical Engineering, Imperial
%College London, South Kensington Campus, London SW7 2BZ, UK }
\affiliation{E. H{\'a}la Laboratory of Thermodynamics, Institute of Chemical Process Fundamentals\\
      of the ASCR, v. v. i., 165 02 Prague 6, Czech Republic\\
Department of Physical Chemistry, Institute of Chemical Technology, Prague, 166 28 Praha 6, Czech Republic}
\author{Andr\'es Santos}
\email{andres@unex.es;
URL:http://www.unex.es/fisteor/andres/}
\affiliation{Departamento de F\'isica, Universidad de Extremadura, E-06071 Badajoz, Spain}

\date{\today}

\begin{abstract}
The one-dimensional penetrable-square-well fluid is studied using
both analytical tools and specialized Monte Carlo simulations.
The model consists of a penetrable core characterized by a finite
repulsive energy combined with a short-range attractive well.
This is a many-body one-dimensional problem, lacking an exact analytical
solution, for which the usual van Hove theorem on the absence of phase transition
does not apply.
We determine a high-penetrability approximation complementing
a similar low-penetrability approximation presented
in previous work. This is shown to be equivalent to the usual
Debye--H\"{u}ckel theory for simple charged fluids for which
the virial and energy routes are { identical}.
The internal thermodynamic consistency with the compressibility
route and the validity of the approximation in describing the radial distribution function is assessed
by a comparison against numerical simulations.
The Fisher--Widom line separating the oscillatory and monotonic large-distance behavior of
the radial distribution function is computed within the high-penetrability approximation
and compared with the opposite regime, thus providing a strong indication of the location of the
line in all possible regimes.
The high-penetrability approximation predicts the existence of a critical point and a spinodal line, but this occurs outside the applicability domain of the theory.
We investigate the possibility of a fluid-fluid transition by Gibbs ensemble Monte Carlo techniques, not finding any evidence of such a transition.  Additional analytical arguments are given to support this claim. Finally, we find a clustering transition when Ruelle's stability criterion is not fulfilled.
The consequences of these findings on the three-dimensional phase diagrams are also discussed.
\end{abstract}

%\pacs{...}
%\keywords{penetrable-square-well, high-penetrability approximation, Monte Carlo simulations, gas-liquid phase transition, Fisher--Widom line.}

\maketitle
%%%%%%%%%%%%%%%%%%%%%%%%%%%%%%%%%%%%%%%%%%%%%%%%%%%%%%%%%%%%%%%%%%%%%%%%%%%%%%%
\section{Introduction}
%%%%%%%%%%%%%%%%%%%%%%%%%%%%%%%%%%%%%%%%%%%%%%%%%%%%%%%%%%%%%%%%%%%%%%%%%%%%%%%
\label{sec:introduction}
Recent advances in chemical synthesis have unveiled more and more the importance of
soft-matter systems, such as dispersions of colloidal particles, polymers, and their
combinations. Besides their practical interest, these new developments have opened up
new theoretical avenues in (at least) two instances. Firstly,  it
is possible to experimentally fine-tune the details of interactions (range, strength, \ldots),
making these systems a unique laboratory for testing highly simplified models
within an effective-interaction approach where the microscopic degrees of freedom are
integrated out in favor of renormalized macroparticle interactions.
Secondly,  they offer the possibility of exploring new types of equilibrium phase
behaviors not present in the simple-fluid paradigm.

As early as in 1989, Marquest and Witten\cite{Marquest89} suggested that the experimentally observed crystallization in some
copolymer micellar systems could be rationalized on the basis of a bounded interaction, that
is, an interaction that  does not diverge at the origin.
Successive theoretical work showed that this class of bounded or ultra-soft potentials naturally arises as
effective interactions between the centers of mass of many soft and flexible macromolecules,
such as polymer chains, dendrimers, star polymers, etc. (see, e.g., Ref.\ \onlinecite{Likos01} for
a reference on the subject). Two well-studied cases belonging to the above class are
the Gaussian core model (GCM) introduced by Stillinger\cite{Stillinger76} and the penetrable-sphere
(PS) model introduced in Refs.\ \onlinecite{Marquest89,Likos98}, whose freezing transition
turns out to display { rather} exotic features with no analogue in the atomistic fluid realm.

In the present paper, we shall consider a close relative of the PS model, first
introduced in Ref.\ \onlinecite{Santos08}, denoted as the penetrable-square-well (PSW) model,
where a short-range attractive tail is added to the PS model just outside the core region.
In the limit of infinite repulsive energy, the PS and PSW models reduce to the usual hard-sphere (HS) and
square-well (SW) models, respectively.

An additional interesting feature common to both PS and PSW, as well as to all bounded potentials, is the fact that,
even when confined to one-dimensional systems, they may  exhibit a non-trivial
phase diagram
due to the penetrability which prevents an exact analytical
solution.

This is because particles cannot be lined up on a line with a well-defined and fixed ordering
in view of the possibility of reciprocal interpenetration (with some positive energy cost), thus lacking an essential
ingredient allowing for an exact solution in the respective hard-core counterparts (HS and SW).
It is then particularly useful to discuss some motivated approximations in the one-dimensional
model which can then be benchmarked against numerical simulations and subsequently exploited in the
much more complex three-dimensional case.

The aim of the present paper is to complete a study on the one-dimensional
PSW model started in Refs.\ \onlinecite{Santos08,Fantoni09}, as well
as the general results presented in Ref.\ \onlinecite{Santos08} which are particularly
relevant in the present context.
In the first paper of the series,\cite{Santos08} we introduced the model
and discussed the range of stability in terms of the attractive versus
repulsive energy scale. We provided, in addition, exact analytical results
in the low-density limit (second order in the radial distribution function
and fourth order in the virial expansion) and a detailed study of the Percus--Yevick
and hypernetted-chain integral equations. These were used in the
following paper\cite{Fantoni09} to propose a low-penetrability approximation at finite density which was then tested against numerical simulation.
This low-penetrability approximation is expected to break down in the opposite regime, namely when temperatures and densities are such that particles easily interpenetrate each other.
In the present paper, we address this latter regime by
proposing a complementary approximation (the high-penetrability approximation) and discussing its range
of validity and the relationship with the low-penetrability regime. Note that a similar matching of the
low- and high-penetrability approximation has already been carried out by two of the present authors
in the framework of the PS model.\cite{Malijevsky06,MYS07}
It turns out that the high-penetrability approximation in the  context of bounded potentials
coincides with the linearized Debye--H\"{u}ckel classical approximation originally introduced in the
framework of the Coulomb potential.\cite{Hansen86} It has been recently shown\cite{Santos09}
that two of the three standard routes to thermodynamics (the energy and the virial routes)
are automatically consistent within the linearized Debye--H\"{u}ckel approximation,
for any potential and dimensionality. This means that a deviation from the
third standard route to thermodynamics (the compressibility route) can be exploited to
assess the degree of reliability of the high-penetrability (or linearized Debye--H\"{u}ckel)
approximation. This is indeed discussed in the present paper, where we also discuss the full hierarchy
of approximations ranging from the full Debye--H\"{u}ckel approximation to the simplest mean-spherical
approximation.

In view of the boundness of the potential, the usual van Hove no-go theorem\cite{VanHove50,Cuesta04} on the absence of phase transitions in certain one-dimensional fluids does
not hold. It is then natural to ask whether a phase transition occurs in the one-dimensional PSW fluid by noting that
the addition of an attractive tail to the pair potential of the PS model extends the question to the
fluid-fluid transition, in addition to the fluid-solid transition possible even within the PS model.
In the present paper we confine our attention to the fluid-fluid case only and discuss this possibility
using both analytical arguments and state-of-the-art numerical simulations.\cite{Panagiotopoulos87,Panagiotopoulos88,Smit89a,Smit89b}
Our results are compatible with the absence of
such a transition, as we shall discuss. This is also supported by recent analytical results\cite{Fantoni10} using a methodology devised for one-dimensional models with long-range interactions.\cite{Kastner08}
We discuss possible reasons for this and a plausible scenario for the three-dimensional case.

Finally, we note that the approach to a critical point is frequently anticipated by
the so-called Fisher--Widom line\cite{Fisher69} marking the borderline
between a region with oscillatory behavior in the long-range domain of the correlation
function (above the Fisher--Widom line) and a region of exponential decay.
We discuss the location of this line within the high-penetrability approximation
and again the matching of this result with that stemming from the low-penetrability
approximation.

The structure of the paper is as follows. We define the
PSW model in Sec.\ \ref{sec:model}. We then construct the high-penetrability approximation in
Sec.\ \ref{sec:HPA} and in Sec. \ref{sec:rHPA} we discuss some
approximations related to it. Section \ref{sec:eos} contains a
discussion on the routes to thermodynamics, as predicted by the
high-penetrability approximation. The structure predicted by the approximation is compared with Monte Carlo data
in Sec.\ \ref{sec:structure}. The Fisher--Widom line and the possibility
of a fluid-fluid transition are discussed in
Sec.\ \ref{sec:FW}. The paper ends with some concluding remarks in
Sec.\ \ref{sec:conclusions}.

%%%%%%%%%%%%%%%%%%%%%%%%%%%%%%%%%%%%%%%%%%%%%%%%%%%%%%%%%%%%%%%%%%%%%%%%%%%%%%%
\section{The penetrable-square-well (PSW) model}
\label{sec:model}
%%%%%%%%%%%%%%%%%%%%%%%%%%%%%%%%%%%%%%%%%%%%%%%%%%%%%%%%%%%%%%%%%%%%%%%%%%%%%%%

The penetrable-square-well (PSW) model is defined by the following
pair potential\cite{Santos08,Fantoni09}
\begin{eqnarray}
\label{model:eq1}
\phi_{\text{PSW}}\left(r\right)=\left\{
\begin{array}{ll}
+\epsilon_r, & r<\sigma,\\
-\epsilon_a, & \sigma<r<\sigma+\Delta,\\
0,           & r>\sigma+\Delta,
\end{array}\right.
\end{eqnarray}
where $\Delta$ is the well width and $\epsilon_r$ and $\epsilon_a$ are two positive constants
accounting for the repulsive and attractive parts of the potential,
respectively. In the following, we shall restrict our analysis to the case $\Delta/\sigma<1$ and $\epsilon_r > 2 \epsilon_a$, where we know the
one-dimensional model to be stable with a well defined thermodynamic limit.\cite{Santos08}
It is shown in Appendix \ref{app:ruelle} that, more generally, the one-dimensional PSW model is guaranteed to be stable
if ${\epsilon_r}>{2(\ell+1)\epsilon_a}$, where $\ell$ is the integer part of $\Delta/\sigma$. For lower values of $\epsilon_r$ the model
may or may not be stable and we will come back to this point in Section \ref{sec:FW}.

An important role in the following is played by the corresponding Mayer function
\begin{equation}
\label{model:eq2}
f_{\text{PSW}}\left(r\right)=\gamma_rf_{\text{HS}}\left(r\right)+\gamma_a\left[\Theta\left(r-\sigma\right)-\Theta\left(r-\sigma-\Delta
\right)\right],
\end{equation}
where $\gamma_r= 1-e^{-\beta\epsilon_r}$ is the parameter measuring the
degree of penetrability varying between 0 (free penetrability) and
1 (impenetrability), while $\gamma_a= e^{\beta\epsilon_a}-1>0$ measures the  strength of the well depth. Here
$\beta=1/k_BT$ with $T$ the temperature and $k_B$ the Boltzmann constant,
$f_{\text{HS}}(r)=-\Theta(\sigma-r)$ is the Mayer function for hard spheres,
and $\Theta(r)$ is the Heaviside step function.

A detailed discussion of the limiting cases of the PSW model can be found in Ref.\ \onlinecite{Santos08}. Here we merely note that
the PSW Mayer function $f_{\text{PSW}}(r)$ is immediately related to the usual SW Mayer function by
\begin{equation}
\label{model:eq3}
f_{\text{PSW}}(r)=\gamma_r f_{\text{SW}}\left(r\right),
\end{equation}
\begin{equation}
\label{model:eq4}
f_{\text{SW}}\left(r\right)= f_{\text{HS}}\left(r\right)+\gamma\left[\Theta\left(r-\sigma\right)-\Theta\left(r-\sigma-\Delta\right)\right]~,
\end{equation}
where we have introduced the ratio $\gamma=\gamma_a/\gamma_r$.
At a given value of $\epsilon_r/\epsilon_a$, $\gamma$ increases quasi-linearly with $e^{\beta\epsilon_a}$, its minimum value $\gamma=\epsilon_a/\epsilon_r$ corresponding to $\beta\epsilon_a\to 0$.

%%%%%%%%%%%%%%%%%%%%%%%%%%%%%%%%%%%%%%%%%%%%%%%%%%%%%%%%%%%%%%%%%%%%%%%%%%%%%%%
\section{A High-Penetrability Approximation (HPA)}
\label{sec:HPA}
%%%%%%%%%%%%%%%%%%%%%%%%%%%%%%%%%%%%%%%%%%%%%%%%%%%%%%%%%%%%%%%%%%%%%%%%%%%%%%%
In Ref.\ \onlinecite{Fantoni09} we discussed a low-penetrability  approximation (LPA) to the PSW model.
Within this approximation, one assumes $1-\gamma_r =e^{-\beta\epsilon_r}\ll 1$ so that the repulsive barrier $\epsilon_r$ is sufficiently higher than the thermal energy $k_BT$,
penetrability is small, and the system is almost a hard-core one. The advantage of this
theoretical scheme is that one can use the general recipe leading to the exact solution for the
 one-dimensional SW problem ---in fact, valid for any potential with a hard core and short-range attractions---
and perform some ad hoc adjustments to ensure that some basic physical conditions on the
radial distribution function $g(r)$ are satisfied. Comparison with Monte Carlo (MC) simulations showed a good behavior of the LPA even for $\beta\epsilon_r=2$ ($1-\gamma_r\simeq 0.14$), provided the density was moderate ($\rho\sigma <0.5$).

The opposite limit $\gamma_r \ll 1$ is also inherently interesting for several reasons.
From a physical viewpoint this amounts to starting from the ideal gas limit $\gamma_r\to 0$ (one of the common reference
systems for simple fluids) and progressively building up interactions by increasing $\gamma_r$.
An additional mathematical advantage stems from the simple observation\cite{Malijevsky06,MYS07,AS04}  that in the (exact) cluster expansion
of $g(r)$ only the dominant chain diagrams need to be retained at all orders, thus leading to
the possibility of an exact summation of those leading contributions. As we shall see { shortly}, this is in fact a procedure known
as the Debye--H\"{u}ckel approximation in the context of charged fluids.\cite{Santos09,Hansen86}

Our main goal is the computation of the cavity function $y(r)\equiv e^{\beta\phi(r)}g(r)$, from which one can immediately compute
the radial distribution function (RDF) $g(r)=y(r)[1+f(r)]$. In the PSW case one then has from
Eq.\ (\ref{model:eq3})
\begin{eqnarray} \label{hpa:eq2}
g(r)&=&\left\{
\begin{array}{ll}
(1-\gamma_r)y(r), & r<1, \\
(1+\gamma_a)y(r), & 1<r<\lambda, \\
y(r),      & r>\lambda.
\end{array}\right.
\end{eqnarray}
where $\lambda=1+\Delta$ and, in conformity with previous work,\cite{Santos08,Fantoni09} we have redefined all lengths in units of $\sigma$
so we set $\sigma=1$ in most of the following equations.

As shown in Ref.\ \onlinecite{AS04} for the PS case, the exact form of the PSW cavity function in the limit $\gamma_r\to 0$ at finite $\gamma=\gamma_a/\gamma_r$ and $\rho\gamma_r$ is
\begin{eqnarray} \label{hpa:eq3}
y\left(r\right)&=&1+\gamma_r w\left(r\right)~,
\end{eqnarray}
where the function $w(r)$ is defined through its Fourier transform
\begin{eqnarray} \label{hpa:eq4}
\widetilde{w}(k)&=&\rho \gamma_r\frac{\widetilde{f}_{\text{SW}}^2(k)}
{1-\rho \gamma_r\widetilde{f}_{\text{SW}}(k)},
\end{eqnarray}
$\widetilde{f}_{\text{SW}}(k)$ being the Fourier transform of ${f}_{\text{SW}}(r)$. Note that in the limit $\gamma_r\to 0$ one has $\gamma_r\approx \beta\epsilon_r$ and $\gamma\approx \epsilon_a/\epsilon_r$.

Generalizing an analogous approximation in the context of the PS model,\cite{Malijevsky06,MYS07} our high-penetrability approximation (HPA)   consists of assuming Eqs.\ (\ref{hpa:eq3}) and (\ref{hpa:eq4}) for finite, but small, values of $\gamma_r$.
It is worth noting that the combination of the two expressions (\ref{hpa:eq3}) and (\ref{hpa:eq4})
defines what is usually referred to, in a different context, as the linearized Debye--H\"{u}ckel (LDH) approximation,\cite{Santos09,Hansen86}
and this will be further elaborated below.

Equations (\ref{hpa:eq3}) and (\ref{hpa:eq4}) { hold} for any dimensionality. In the specific one-dimensional case, and taking into account Eq.\ \eqref{model:eq4}, we have
\begin{eqnarray}
\nonumber
\widetilde{f}_{\text{SW}}\left(k\right)&=&2\int_0^\infty dr \cos(kr) f_{\text{SW}}(r)\\ \label{hpa:eq1}
&=&-\frac{2}{k}\left[\left(1+\gamma\right){\sin k}-
\gamma{\sin\lambda k}\right]~.
\end{eqnarray}
The function $w(r)$ can be numerically evaluated in real space by Fourier inversion as
\begin{equation}
w(r)=\frac{\rho \gamma_r}{\pi}\int_0^\infty dk\, \cos(kr)\frac{\widetilde{f}_{\text{SW}}^2(k)}
{1-\rho \gamma_r\widetilde{f}_{\text{SW}}(k)}.
\label{w(r)}
\end{equation}
An explicit expression for the density expansion of $w(r)$ within the HPA is reported in Eqs.\ (\ref{appa:eq6}) and (\ref{appa:eq7}) of Appendix \ref{app:appa}, where the radius of convergence of the expansion is also analyzed.

From Eqs.\ (\ref{model:eq3}), (\ref{hpa:eq2}), and (\ref{hpa:eq3}) the
total correlation function, $h(r)=g(r)-1$, within the HPA is easily obtained as
\begin{eqnarray} \label{hpa:eq5}
h\left(r\right)&=&\gamma_rw\left(r\right)\left[1+\gamma_rf_{\SW}\left(r\right)\right]+\gamma_rf_{\SW}\left(r\right)~,
\end{eqnarray}
or in Fourier space,
\begin{eqnarray} \label{hpa:eq6}
\widetilde{h}\left(k\right)&=&
\gamma_r\frac{\widetilde{f}_{\text{SW}}\left(k\right)}{1-\rho\gamma_r\widetilde{f}_{\text{SW}}\left(k\right)}\nonumber\\
&&+\rho\gamma_r^3\int_{-\infty}^{\infty}\frac{dk^\prime}{2\pi}\,
\frac{
\widetilde{f}^2_{\text{SW}}(k^\prime)\widetilde{f}_{\text{SW}}(\left\vert k-k^\prime\right\vert)}{1-\rho\gamma_r\widetilde{f}_{\text{SW}}\left(k^\prime\right)}.
\end{eqnarray}
From this equation it is straightforward to get the structure factor
\begin{eqnarray}
 S(k)&=&1+\rho\widetilde{h}(k)\nonumber\\
 &=&\frac{1}{1-\rho\gamma_r\widetilde{f}_{\text{SW}}\left(k\right)}\nonumber\\
&&+\rho^2\gamma_r^3\int_{-\infty}^{\infty}\frac{dk^\prime}{2\pi}\,
\frac{
\widetilde{f}^2_{\text{SW}}(k^\prime)\widetilde{f}_{\text{SW}}(\left\vert k-k^\prime\right\vert)}{1-\rho\gamma_r\widetilde{f}_{\text{SW}}\left(k^\prime\right)}
\end{eqnarray}
and the Fourier transform of the direct
correlation function $\widetilde{c}(k)=\widetilde{h}(k)/S(k)$. The zero wavenumber value of the structure factor is
\begin{equation}
 S(0)=\frac{1}{1+2\rho\gamma_r(1-\gamma\Delta)}
+\rho^2\gamma_r^3\int_{0}^{\infty}\frac{dk}{\pi}\,
\frac{
\widetilde{f}^3_{\text{SW}}(k)}{1-\rho\gamma_r\widetilde{f}_{\text{SW}}\left(k\right)},
\label{S(0)}
\end{equation}
where we have taken into account that $\widetilde{f}_{\text{SW}}\left(0\right)=-2(1-\gamma\Delta)$.
This completes the calculation of the correlation functions within the HPA.

%%%%%%%%%%%%%%%%%%%%%%%%%%%%%%%%%%%%%%%%%%%%%%%%%%%%%%%%%%%%%%%%%%%%%%%%%%%%%%%
\section{Approximations related to the HPA}
\label{sec:rHPA}
%%%%%%%%%%%%%%%%%%%%%%%%%%%%%%%%%%%%%%%%%%%%%%%%%%%%%%%%%%%%%%%%%%%%%%%%%%%%%%%
As anticipated in Section \ref{sec:HPA}, the HPA is the exact equivalent to the well-known LDH approximation,
which is widely used in the context of charged simple fluids.\cite{Hansen86} The latter is actually
an intermediate step of a hierarchy of successive approximations ranging from the simplest mean-spherical approximation
(MSA) to the full non-linear version of the Debye--H\"{uckel} approximation (see Ref.\ [11] in Ref.\ \onlinecite{Santos09}
for a discussion on this point). For the PSW model ---and more generally for any bounded potential--- the LDH approximation (the HPA in the present language)
is particularly relevant in view of the fact that one can make $f_{\text{PSW}}(r)$ arbitrarily small by letting $\gamma_r \to 0$, thus justifying
the approximation of neglecting non-chain diagrams.
It is then interesting to check the performance of the other approximations included in the aforementioned class, which will be translated in the
present context for simplicity.

On top of the hierarchy of approximations there is the non-linear HPA (nlHPA)
\begin{eqnarray}
\label{rHPA:eq1}
y\left(r\right)&=&e^{\gamma_r w\left(r\right)}~,
\end{eqnarray}
which is equivalent to the non-linear Debye--H\"{uckel} approximation, as remarked. The HPA, Eq.\ \eqref{hpa:eq3},  is obtained upon linearizing
the exponential, an approximation valid again in the limit $\gamma_r\ll 1$.
An additional approximation ---denoted here as the modified HPA (mHPA)--- can be considered with the help of Eq.\ (\ref{hpa:eq5})
by neglecting the quadratic term  in $\gamma_r$. This yields
\begin{eqnarray}
\label{rHPA:eq2}
h\left(r\right)&=&\gamma_r\left[w(r)+f_{\text{SW}}\left(r\right)\right]~.
\end{eqnarray}
This is equivalent to keeping only the first term on the right-hand side of Eq.\ \eqref{hpa:eq6}, which implies $\widetilde{c}\left(k\right)=\gamma_r\widetilde{f}_{\text{SW}}\left(k\right)$ or, in real space,
\begin{eqnarray}
\label{rHPA:eq3}
c\left(r\right)&=&f_{\text{PSW}}\left(r\right)~.
\end{eqnarray}
The lowest rank in the hierarchy is occupied by the MSA,
which is obtained from Eq.\ (\ref{rHPA:eq3}) upon linearization of the Mayer function
$f_{\text{PSW}}(r)$,
\begin{eqnarray}
\label{rHPA:eq4}
c\left(r\right)&=&-\beta\phi_{\text{PSW}}\left(r\right)~.
\end{eqnarray}

Since $w(r)$ is a convolution, it must be continuous at $r=1$ and $r=\lambda$. It
follows that the approximations with a continuous cavity function at
$r=1$ and $r=\lambda$ are nlHPA and HPA.
For instance, in the mHPA \eqref{rHPA:eq2}  the cavity function is $y(r)=1+\gamma_r w(r)/[1+\gamma_r f_\SW(r)]$, so that one has $y(1^-)-y(1^+)=\gamma_r^2 w(1)(1+\gamma)(1-\gamma_r)^{-1}(1+\gamma\gamma_r)^{-1}$ and $y(\lambda^+)-y(\lambda^-)=\gamma_r^2 w(\lambda)\gamma(1+\gamma\gamma_r)^{-1}$.

 It has been shown in Ref.\ \onlinecite{Santos09} that the virial and energy routes to thermodynamics
(to be discussed below) are consistent one another within the HPA, for any potential and any dimensionality.
A similar statement holds true for soft potentials within the MSA.\cite{Santos07}
This clearly includes the PSW potential in both cases.

It is interesting to make contact with previous work carried out by Likos et al. \cite{Likos01b,Likos07} on a general class
of unbounded potentials which are free of attractive parts, thus resulting particularly useful in the
context of the fluid-solid transition. In Ref. \onlinecite{Likos01b}, the MSA given in Eq.\ (\ref{rHPA:eq4}) along with
the spinodal instability to be discussed in detail in Section \ref{sec:FW}, have been introduced
for a general class of unbounded potentials including the PS as a particular case. This has been further
elaborated and extended to include the GCM in Ref. \onlinecite{Likos07}. In both cases, the authors discuss directly
the three-dimensional case, so that a direct comparison with the present work cannot be drawn at the present stage,
but they also provide a detailed discussion of various approximations, within the general framework of
density functional theory, that provide a unified framework where even the present model could be included.

%%%%%%%%%%%%%%%%%%%%%%%%%%%%%%%%%%%%%%%%%%%%%%%%%%%%%%%%%%%%%%%%%%%%%%%%%%%%%%%%%%%%%%%%%%%%%%%%%%%
\section{Equation of state }
\label{sec:eos}
%%%%%%%%%%%%%%%%%%%%%%%%%%%%%%%%%%%%%%%%%%%%%%%%%%%%%%%%%%%%%%%%%%%%%%%%%%%%%
Given an approximate solution of a fluid model there are several
routes to the equation of state which, in general, give different results. The
most common are three:\cite{Hansen86} the virial, the compressibility, and the energy
route. The consistency of the outcome of these different routes
can be regarded as an assessment on the soundness of the approximation.
For some particular approximations it may also happen that the consistency of two of the three routes
are automatically enforced (see Ref.\ \onlinecite{Santos09} and references therein for a detailed discussion on this point).
This is the case of the HPA, where the virial and energy routes coincide, as anticipated.
Hence, the consistency with the compressibility route will provide a rough estimate of the regime of validity
of the HPA within a phase diagram for the PSW potential.

Let us briefly recall\cite{Hansen86} the methods to compute the compressibility factor, $Z=\beta p/\rho$, associated with the
three different routes.
The virial route is defined by
\begin{eqnarray}\label{eos:eq1}
Z&=&1-\rho \beta\int_{0}^{\infty} dr \, r y\left(r\right) e^{-\beta \phi\left(r\right)}  \phi^{\prime}\left(r\right),
\end{eqnarray}
which, using standard manipulations,\cite{Hansen86} yields
\begin{eqnarray}
\label{eos:eqs2}
Z&=&1+\rho\gamma_r\left[\left(1+\gamma\right)y\left(1\right)-\gamma\lambda y\left(\lambda\right)\right].
\end{eqnarray}
Thus the problem is reduced to the computation of the cavity function $y(r)$ which, in the
present context, follows from Eqs.\ (\ref{hpa:eq3}) and (\ref{w(r)}).

Next, we consider the compressibility route,
\begin{eqnarray}
\label{eos:eq3}
Z&=&\frac{1}{\rho}\int_0^\rho\frac{d\rho'}{S(0)}~,
\end{eqnarray}
where the integral can be readily evaluated with the help of
Eq.\ (\ref{S(0)}).

Regarding the energy route, we start from the internal energy per
particle
\begin{eqnarray} \label{eos:eq5}
u&=&\frac{1}{2\beta}+\epsilon_r\left(1-\gamma_r\right)\rho\int_0^1 dr\,y\left(r\right)\nonumber\\
&&-
\epsilon_a\left(1+\gamma_a\right)\rho\int_1^{\lambda} dr\,y\left(r\right).
\end{eqnarray}
In the above equation, the expressions given by Eqs.\ (\ref{model:eq1}) and (\ref{hpa:eq2})
have been used.
In order to obtain $\beta p$ from $u$ we exploit the following
standard thermodynamic identity\cite{Santos09}
\begin{eqnarray}\label{eos:eq6}
\rho^2\left(\frac{\partial u}{\partial\rho}\right)_\beta&=&
\left(\frac{\partial \beta p}{\partial\beta}\right)_\rho~,
\end{eqnarray}
thus leading to
\begin{eqnarray}
\label{eos:eq7}
Z&=&1+\rho\int_{0}^{\beta} d\beta'\,
\left(\frac{\partial u}{\partial\rho}\right)_{\beta'}~.
\end{eqnarray}
We have used the exact  consistency between the virial and energy routes within the HPA as a test of the numerical calculations.

\begin{figure}
\includegraphics[width=\columnwidth]{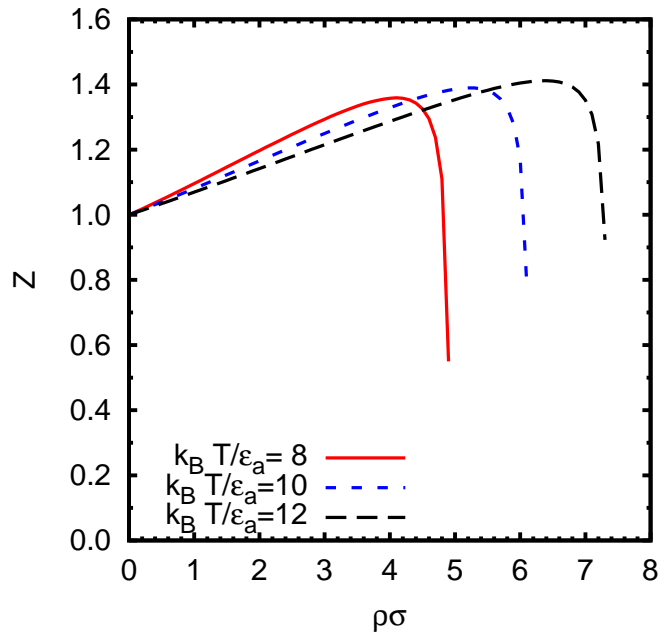}
\caption{Plot of the compressibility factor $Z=\beta p/\rho$ as computed from the virial(-energy) route to the pressure for
$\epsilon_r/\epsilon_a=2$ and $\Delta/\sigma=1$. Results for three different reduced temperatures $k_B T/\epsilon_a=8,10,12$
are displayed.
\label{fig:fig1}}
\end{figure}

Figure \ref{fig:fig1} depicts the results of the virial ---and hence energy, as remarked--- route under the condition
$\epsilon_r/\epsilon_a=2$, which constitutes the borderline range of stability of the one-dimensional PSW model with $\Delta/\sigma<1$.\cite{Santos08}
Under this demanding condition, we have considered three reduced temperatures from $k_B T /\epsilon_a=8$ to  $k_B T /\epsilon_a=12$,
whereas the width of the well has been fixed to the value $\Delta/\sigma=1$.
We remark that $\rho \sigma$ is not limited in values from above due to the boundness of the potential.
The clear downturn of all three curves for sufficiently large reduced density $\rho \sigma$  is
a consequence of the existence of a maximum density $\rho_{\max}$ [see Eq.\ \eqref{hpa:eq9}],
beyond which the HPA breaks down, as described at the end of Appendix \ref{app:appa}. In particular, the values of the maximum density for $\epsilon_r/\epsilon_a=2$ and $\Delta/\sigma=1$ are  $\rho_{\max}\sigma=4.94$, $6.15$, and $7.36$ at $k_BT/\epsilon_a=8$, $10$, and $12$, respectively, in agreement with Fig.\ \ref{fig:fig1}.

\begin{figure}
\includegraphics[width=\columnwidth]{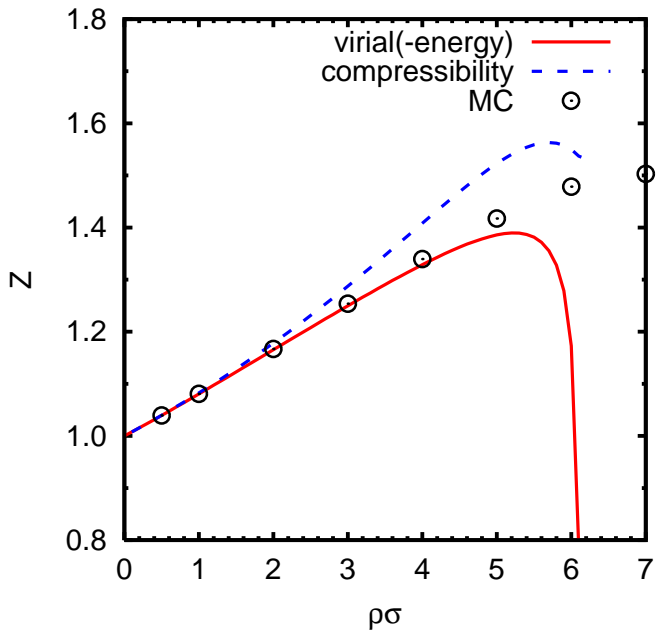}
\caption{Comparison between the compressibility factor $Z=\beta p/\sigma$ from the virial(-energy) and the
compressibility route to the pressure. The circles represent MC simulation results. Chosen parameters are $\epsilon_r/\epsilon_a=2$, $\Delta/\sigma =1$, and $k_B
T/\epsilon_a=10$.
\label{fig:fig2}}
\end{figure}

We compare in Fig.\ \ref{fig:fig2} the results from the virial(-energy) and the compressibility
routes with MC simulations\cite{note1} for an intermediate value of the reduced temperature ($k_B T /\epsilon_a=10$)
and other parameters as before.
Rather interestingly, the virial(-energy) route appears to reproduce rather well the numerical simulation
results up to the region where the artificial downward behavior shows up, whereas the compressibility route begins to deviate
for densities $\rho \sigma>3$.

\begin{figure}
\includegraphics[width=\columnwidth]{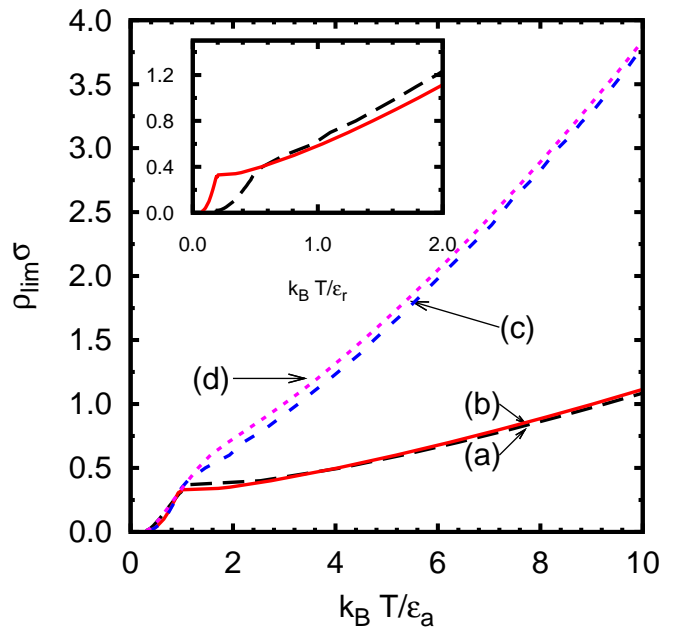}
\caption{Rough estimate of the region of reliability for the HPA,
on the basis of the consistency between the virial(-energy) and compressibility routes,  in
the reduced density $\rho \sigma$ versus reduced temperature $k_B T/\epsilon_a$ plane.
Here the curves (a)--(d) correspond to $(\epsilon_r/\epsilon_a,\Delta/\sigma)=(5,0.5)$, $(5,1)$, $(2,1)$, and $(2,0.5)$, respectively. The region below each curve represents
states where the relative deviation between the virial(-energy) route and the compressibility
one is smaller than $5\%$ and hence regarded as reliable. The inset shows the curves in the $\rho \sigma$ versus  $k_B T/\epsilon_r$ plane for $(\epsilon_r/\epsilon_a,\Delta/\sigma)=(2,1)$ (dashed line) and $(5,1)$ (solid line).
\label{fig:fig3}}
\end{figure}

As the temperature increases the HPA theory clearly remains a good
approximation for a larger range of densities. We can naturally measure this by the requirement that virial(-energy)
and compressibility routes are consistent within a few percent. This is indeed shown in Fig.\ \ref{fig:fig3},
where we depict a transition line separating a ``reliable'' from an ``unreliable'' regime, as measured by the
relative deviation of the two routes (here taken to be $5\%$), for four choices of the model parameters: $(\epsilon_r/\epsilon_a,\Delta/\sigma)=(2,1)$, $(2,0.5)$, $(5,1)$, and $(5,0.5)$. The value $\Delta/\sigma=0.5$ is frequently used in the
SW counterpart.\cite{Vega95} We observe that the region $0\leq \rho\leq \rho_\text{lim}(T)$ where the HPA is reliable is hardly dependent on $\Delta/\sigma$ [compare curves (a)\&(b) and (c)\&(d) in Fig.\ \ref{fig:fig3}]. On the other hand, at  given values of $\Delta/\sigma$ and $k_BT/\epsilon_a$, the range $0\leq \rho\leq \rho_\text{lim}(T)$ decreases with increasing $\epsilon_r/\epsilon_a$ [compare
curves (a)\&(d) and (b)\&(c) in Fig.\ \ref{fig:fig3}], as expected. However, this effect is much less important if the increase of $\epsilon_r/\epsilon_a$ takes place at fixed $k_BT/\epsilon_r$ (see inset of Fig.\ \ref{fig:fig3}). It is interesting to note that, as illustrated by Fig.\ \ref{fig:fig2}, the HPA virial route keeps being reliable up to a { certain} density higher than $\rho_\text{lim}$.

\begin{figure}
\includegraphics[width=\columnwidth]{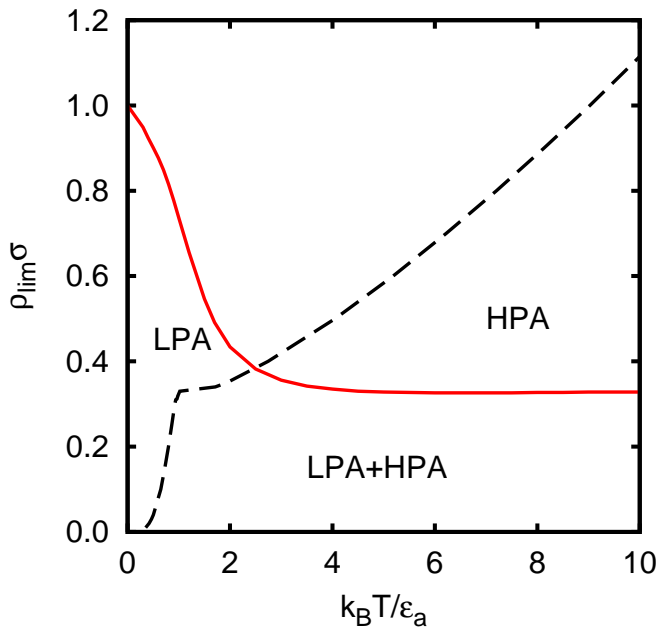}
\caption{Regions of reliability for the LPA and the HPA  in
the reduced density $\rho \sigma$ versus reduced temperature $k_B T/\epsilon_a$ plane.
Here $\epsilon_r/\epsilon_a$=5 and $\Delta/\sigma=1$. The labels LPA, HPA, and LPA+HPA indicate the regions where only the LPA, only the HPA, or both approximations are reliable, respectively.
\label{fig:fig3b}}
\end{figure}

As said above, in Ref.\ \onlinecite{Fantoni09} we introduced a LPA  that was accurate for states where the penetrability effects were low or moderate. { The} LPA is complemented by the HPA presented in this paper. It is { then} interesting to compare the regions where each approximation can be considered reliable according to the same criterion as in Fig.\ \ref{fig:fig3}. This is shown in { Fig.\ }  \ref{fig:fig3b} for the case $\epsilon_r/\epsilon_a=5$ and $\Delta/\sigma=1$. The two transition lines split the plane into four regions: a region where only the LPA is reliable, a region where only the HPA is reliable, a region where both approximations are reliable (and provide equivalent results), and a region where none of them is sufficiently good. The latter region shrinks as { $\epsilon_r/\epsilon_a$} decreases (thanks to the HPA) or
{ $\Delta/\sigma$} decreases (thanks to the LPA).

%%%%%%%%%%%%%%%%%%%%%%%%%%%%%%%%%%%%%%%%%%%%%%%%%%%%%%%%%%%%%%%%%%%%%%%%%%%%%%%
\section{Structure}
\label{sec:structure}
%%%%%%%%%%%%%%%%%%%%%%%%%%%%%%%%%%%%%%%%%%%%%%%%%%%%%%%%%%%%%%%%%%%%%%%%%%%%%%%

\begin{figure}
\includegraphics[width=\columnwidth]{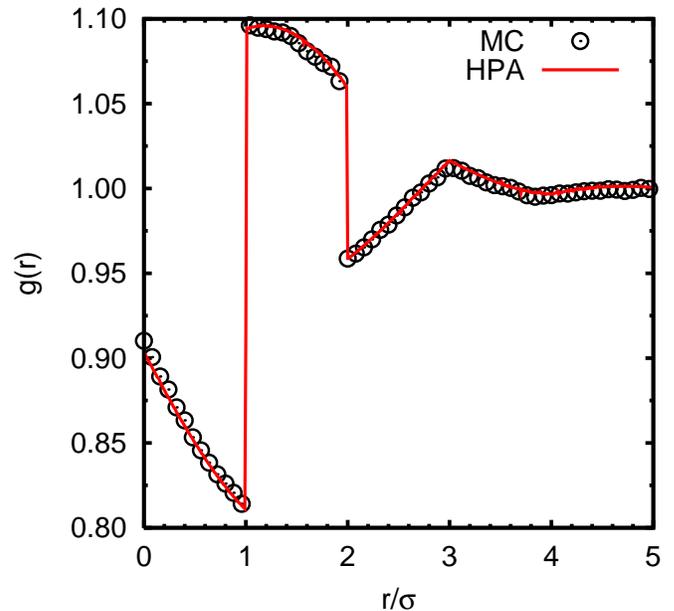}
\caption{Results for the RDF $g(r)$ as a function of $r/\sigma$ with $\epsilon_r
/\epsilon_a=2$,
$\Delta/\sigma=1$, $k_BT/\epsilon_a=10$,  and $\rho \sigma=1.5$. Predictions from the HPA (solid
line) are compared with MC results (circles).
\label{fig:fig4}}
\end{figure}

As an additional test of the soundness of the HPA, we also study the
RDF $g(r)=h(r)+1$, which can easily be obtained
from Eqs.\ \eqref{hpa:eq2} and \eqref{hpa:eq3}, or equivalently from Eq.\ (\ref{hpa:eq5}), once  the auxiliary function $w(r)$ has been determined.
For a sufficiently high temperature (and hence high penetrability), the HPA is clearly well
performing, as can be inferred from Fig.\ \ref{fig:fig4}, when compared with standard NVT MC results. Here we have
considered the same parameters as in the preceding section ($\epsilon_r/\epsilon_a=2$ and $\Delta/\sigma=1$) at
a corresponding high-temperature value $k_BT/\epsilon_a=10$ and a density $\rho\sigma=1.5$ where overlappings are unavoidable.  Under these conditions,
there is no visible difference among the various approximations considered in Section \ref{sec:rHPA}. The excellent performance of the HPA  observed in Fig.\ \ref{fig:fig4} agrees with the reliability criterion of Fig.\ \ref{fig:fig3} since the state $k_BT/\epsilon_a=10$ and $\rho\sigma=1.5$ is well below the curve (c) corresponding to $\epsilon_r/\epsilon_a=2$ and $\Delta/\sigma=1$.

\begin{figure}
\includegraphics[width=\columnwidth]{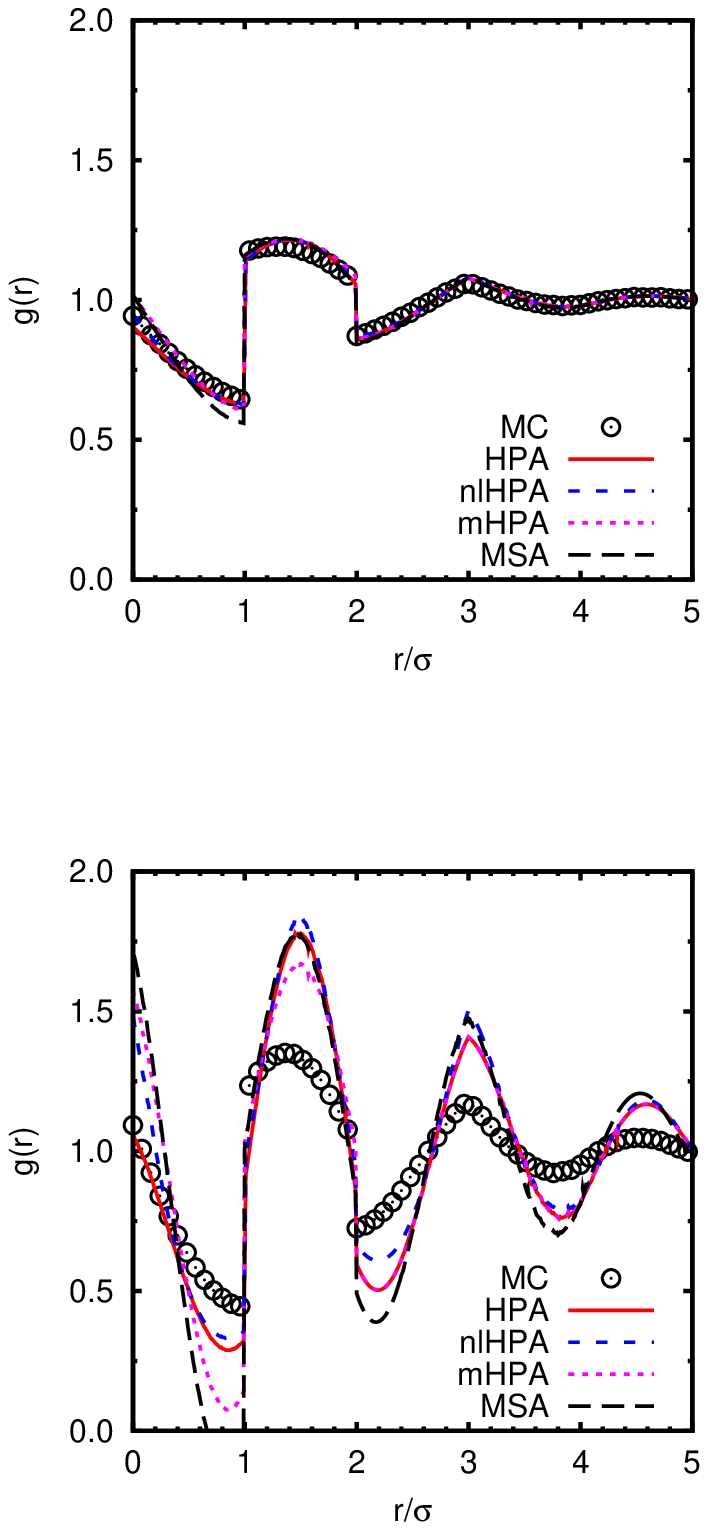}
\caption{Comparison of different approximations in the results for the RDF $g(r)$ vs $r/\sigma$.
Chosen parameters are $\epsilon_r
/\epsilon_a=2$, $\Delta/\sigma=1$, and $\rho \sigma=1.5$. MC results (circles)
are compared with the HPA (solid line), the nlHPA (dashed line), the mHPA (short dashed line), and the MSA (long dashed line). The two different panels refer to different reduced temperatures: $k_B T/\epsilon_a=5$ (top panel)
and $k_B T/\epsilon_a=3$ (bottom panel). Note that both panels are drawn within the same scale.
\label{fig:fig5}}
\end{figure}

As we cool down, significant differences among various approximations (HPA, nlHPA, mHPA, and MSA) begin to appear, as depicted in Fig.\ \ref{fig:fig5}, where
results corresponding to temperatures $k_BT/\epsilon_a=5$ (top panel) and $k_BT/\epsilon_a=3$ (bottom panel) are reported within the same scale.
The state $k_BT/\epsilon_a=5$ and $\rho\sigma=1.5$ is still lying in the reliable region of Fig.\ \ref{fig:fig3}, but close to the boundary line (c), while the state { $k_BT/\epsilon_a=3$} and $\rho\sigma=1.5$ is clearly outside that region. In the case $k_BT/\epsilon_a=5$ the HPA and its three variants are practically indistinguishable, except  in the
region $0<r<\sigma$, which is very important to describe the correct thermodynamic behavior, where the best agreement with MC data corresponds to the nlHPA, followed by the HPA. The two approximations that do not preserve the continuity of the cavity functions (mHPA and MSA) overestimate the jump
at $r=\sigma$. In the lower temperature case $k_BT/\epsilon_a=3$ all the approximations { overestimate} the oscillations of the RDF. Interestingly, the HPA captures quite well the values of $g(r)$ near the origin. The worst overall behavior corresponds again to the MSA, which even predicts negative values of $g(r)$ for $r/\sigma\lesssim 1$.

\begin{figure}
\includegraphics[width=\columnwidth]{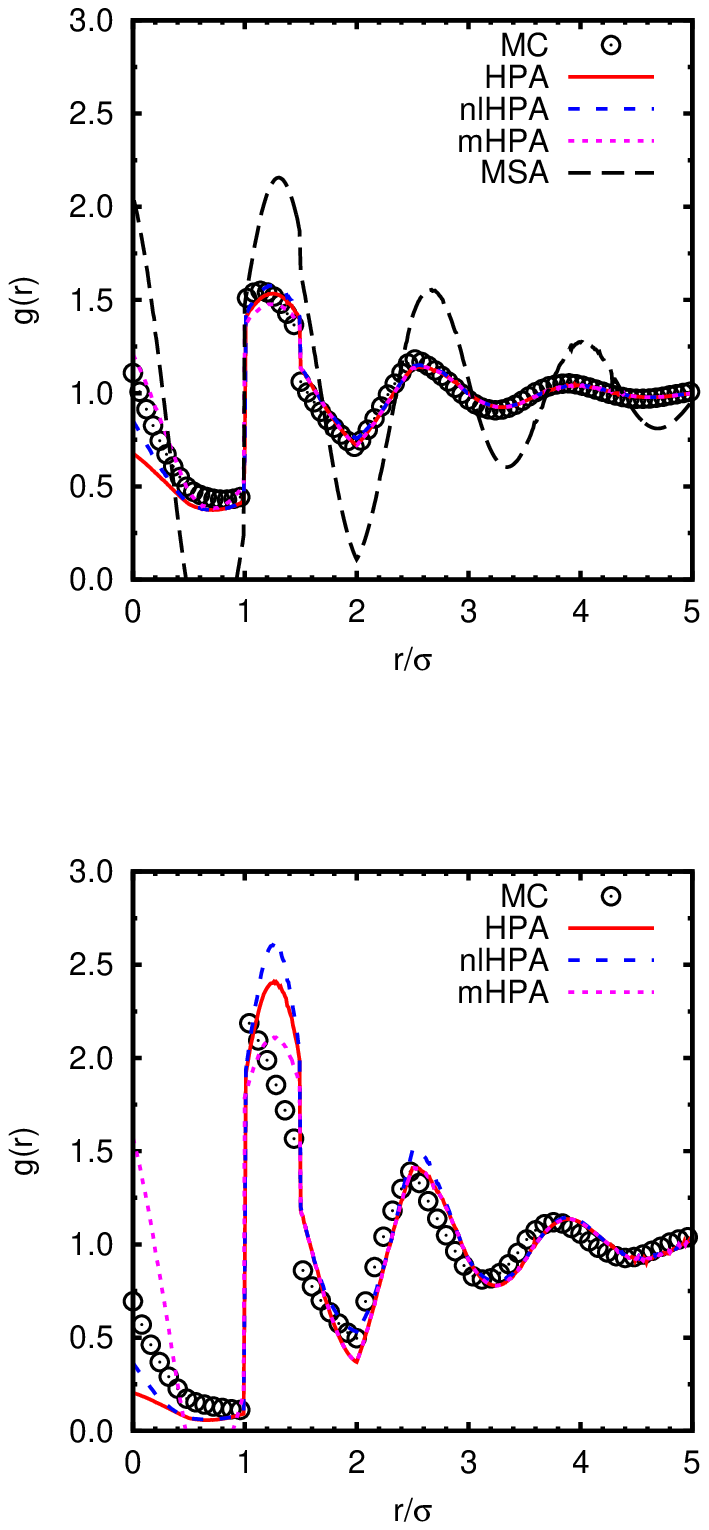}
\caption{An additional comparison of different approximations in the results for the RDF $g(r)$ vs $r/\sigma$.
Here fixed parameters are $\epsilon_r
/\epsilon_a=5$ and $\Delta/\sigma=0.5$. MC results (circles) are compared with the HPA (solid line), the nlHPA (dashed line), the mHPA
(short dashed line),
and the MSA (long dashed line). The top panel  refers to the state $k_B T/\epsilon_a=5$ and $\rho \sigma=1.5$, whereas
the bottom panel  refers to  $k_B T/\epsilon_a=2$ and $\rho \sigma=0.8$. In the bottom panel the MSA is not depicted.
{ Again both panels are on the same scale.}
\label{fig:fig6}}
\end{figure}

\begin{figure}
\includegraphics[width=\columnwidth]{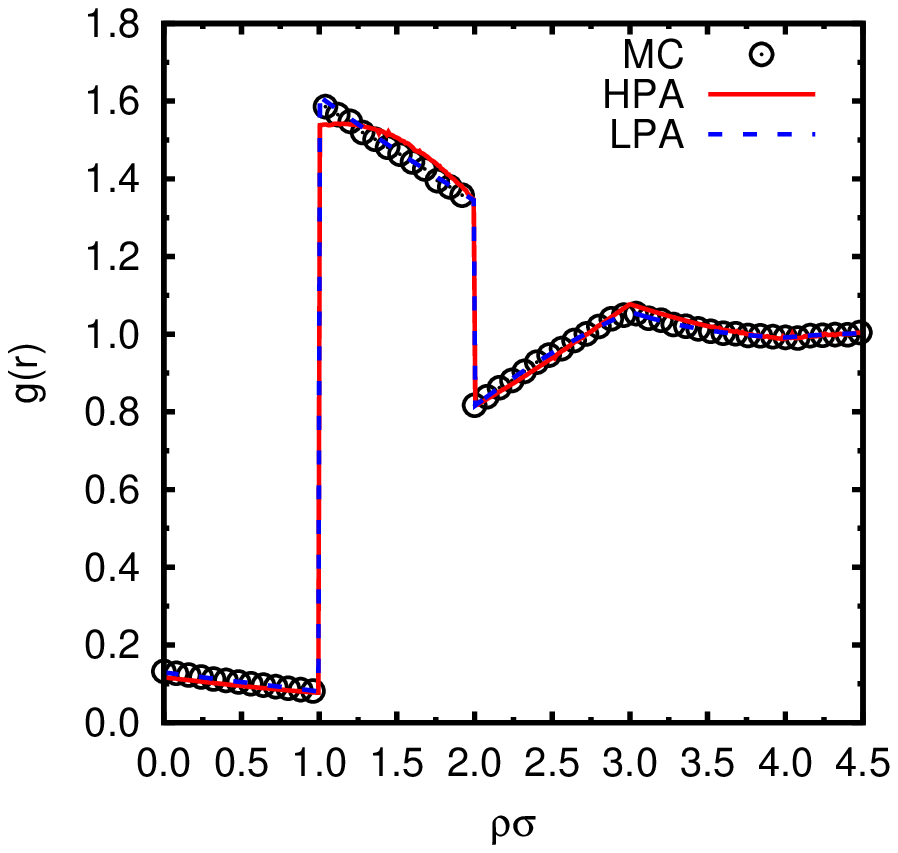}
\caption{Comparison of different approximations in the results for the RDF $g(r)$ vs $r/\sigma$.
Chosen parameters are $\epsilon_r/\epsilon_a=5$, $\Delta/\sigma=1$, $k_BT/\epsilon_a=2$, and $\rho\sigma=0.2$. MC results (circles)
are compared with the HPA (solid line) and the LPA (dashed line).
\label{fig:fig6b}}
\end{figure}

Additional insights can be obtained by decreasing the range of interactions, in
close analogy with what we  considered in previous work for the complementary
LPA.\cite{Fantoni09} This is reported in
Fig.\ \ref{fig:fig6} for the case $\epsilon_r/\epsilon_a=5$ and $\Delta/\sigma=0.5$ with $(k_BT/\epsilon_a,\rho\sigma)=(5,1.5)$ (top panel) and $(k_BT/\epsilon_a,\rho\sigma)=(2,0.8)$ (bottom panel).  Again we stress that the same scale is used for both panels
in order to emphasize the effect of lowering the temperature.
Clearly this is a more demanding situation. In fact, both states are above curve (a) of Fig.\ \ref{fig:fig3} and thus outside  the corresponding reliability region. Therefore,
 clear deviations from MC results
appear in all considered approximations, especially in the lower temperature case $k_BT/\epsilon_a=2$  (bottom panel). Yet,
the HPA is still a reasonably good approximation that follows the main qualitative features of the correct $g(r)$. In the higher temperature case $k_BT/\epsilon_a=5$ the only noticeable limitations of the HPA practically take place near the origin, this deficiency being largely corrected by the mHPA.

To conclude this section, it is worthwhile comparing the two complementary approaches HPA and LPA at a case where both are expected to be reliable, according to the diagram of Fig.\ \ref{fig:fig3b}. This is done in Fig.\ \ref{fig:fig6b} for $\epsilon_r/\epsilon_a=5$, $\Delta/\sigma=1$, $k_BT/\epsilon_a=2$, and $\rho\sigma=0.2$. We observe that both approximations agree well each other and with MC data, except in the region $1<r/\sigma<\Delta/\sigma$, where the HPA RDF presents an artificial curvature.

%%%%%%%%%%%%%%%%%%%%%%%%%%%%%%%%%%%%%%%%%%%%%%%%%%%%%%%%%%%%%%%%%%%%%%%%%%%%%%%
\section{Fisher--Widom line and fluid-fluid transition}
\label{sec:FW}
%%%%%%%%%%%%%%%%%%%%%%%%%%%%%%%%%%%%%%%%%%%%%%%%%%%%%%%%%%%%%%%%%%%%%%%%%%%%%%%
We now turn to an interesting point raised in previous work,\cite{Fantoni09} namely the question of whether the model can display
a phase transition in spite of its one-dimensional character.

The existence of general theorems { ---all essentially based on the original van Hove's result\cite{VanHove50}--- }
on the absence of phase transitions
for a large class of one-dimensional models with short-range interactions { is } well established.\cite{Cuesta04} PSW and PS models, however, do not belong to the class for which these general theorems
hold. This is because boundness allows multiple partial (or even total) overlapping
at some energy cost, thus rendering { invalid the arguments used in the aforementioned theorems.}

On the other hand, none of these theorems provide a general guideline to
understand whether a one-dimensional model may or may not display a non-trivial
phase transition, and one has then to rely on the specificity of each model.

As discussed in our previous work,\cite{Fantoni09} it is instructive to
first address the simpler question of the location of the Fisher--Widom (FW) line.
This is a line separating two different regimes for the
large-distance behavior of the RDF $g(r)$ in the presence of competing repulsive/attractive interactions.\cite{Fisher69}
The rationale behind the FW line is that on approaching the critical points where attractions
become more and more effective, the behavior of correlation functions must switch from oscillatory (characteristic
of repulsive interactions) to exponential with a well defined correlation length $\xi$.
In the previous work, we analyzed the location of this line for PSW within the LPA. Here we extend this
analysis to the HPA regime and discuss the compatibility of the two results.

Let us first briefly recall the main points of the analysis, referring to Ref.\ \onlinecite{Fantoni09}
for details.
From Eq.\ (\ref{hpa:eq2}) we note that the asymptotic behavior of $g(r)$ is the same as that of $y(r)$. In view of Eq.\ (\ref{hpa:eq3}), this is hence related to $w(r)$, whose asymptotic behavior is governed by the pair of
conjugate poles of $\widetilde{w}(k)$ with an imaginary part closest to
the origin. If the real part of the pair is zero, the decay is
monotonic and oscillatory otherwise.

According to Eq.\ \eqref{hpa:eq4}, the poles of $\widetilde{w}(k)$ are given by
\begin{eqnarray}
\label{fw:eq1}
\rho\gamma_r\widetilde{f}_{\text{SW}}\left(k\right)&=&1.
\end{eqnarray}
Let $k=\pm i x$ be the imaginary pole and $k=\pm i(x^\prime\pm iy)$ be
the pole with the imaginary part closest to the origin. The FW line is
determined by the condition $x=x^\prime$. This gives, at a given temperature, three equations in the
three unknowns $x$, $y$, and $\rho$.\cite{Fantoni09} More specifically, after some algebra, one gets
\begin{eqnarray}
\frac{y}{x}\sinh x&\sinh (\lambda x)&\left[\cos y-\cos(\lambda y)
\right]=
\sinh  x\cosh(\lambda x)\nonumber\\
&&\times\sin(\lambda y)(\cos y-1)-\sinh (\lambda x)\cosh x\nonumber\\
&&\times\sin  y\left[\cos (\lambda y)-1\right] ,
\label{fw:eq2}
\end{eqnarray}
\begin{equation}
\gamma=\left[\frac{\sinh (\lambda x)}{\sinh  x}\frac{\cos (\lambda y)-1}{\cos y-1}-1\right]^{-1},
\label{fw:eq3}
\end{equation}
\begin{equation}
\rho=\frac{1}{2\gamma_r}\frac{x}{\gamma\sinh (\lambda x)-(1+\gamma)\sinh x}.
\label{fw:eq4}
\end{equation}
The inverse of the parameter $x$ represents the correlation length $\xi=1/x$. From a practical point of view it is { more} convenient to use $x$ rather than $k_BT/\epsilon_a$ as a free parameter to construct the FW line. In that way, Eq.\ \eqref{fw:eq2} becomes a transcendental equation that gives $y$ as a function of $x$;  once $y(x)$ is known, the solution to Eq.\ \eqref{fw:eq3} gives $k_BT/\epsilon_a$ as a function of $x$;  finally, insertion of $y(x)$ and $k_BT(x)/\epsilon_a$ into Eq.\ \eqref{fw:eq4} provides $\rho(x)$. The corresponding values of the pressure are obtained from either Eq.\ \eqref{eos:eqs2} (virial-energy route) or Eq.\ \eqref{eos:eq3} (compressibility route).

We observe that $T$ decreases as $x$ decreases, until a \emph{critical} value $T_c$ is found in the limit $x\to 0$. In that limit, Eqs.\ \eqref{fw:eq2}--\eqref{fw:eq4} { simplify to}
\begin{eqnarray}
\lambda{y_c}\left[\cos y_c-\cos(\lambda y_c)\right]&=&
 \sin(\lambda y_c)(\cos y_c-1)-\lambda\sin  y_c\nonumber\\
 &&\times\left[\cos (\lambda y_c)-1\right] ,
\label{fw:eq5}
\end{eqnarray}
\begin{equation}
\gamma_c=\left[\lambda\frac{\cos (\lambda y_c)-1}{\cos y_c-1}-1\right]^{-1},
\label{fw:eq6}
\end{equation}
\begin{equation}
\rho_c=\frac{1}{2\gamma_r}\frac{1}{\gamma_c\Delta-1}.
\label{fw:eq7}
\end{equation}
At  the critical point $T=T_c$ and $\rho=\rho_c$, one has $x=0$ or, equivalently, $\xi\to\infty$. Therefore, at this point $w(r)$ does not decay  for long distances and in Fourier space one has $\widetilde{w}(k)\sim k^{-2}$ and $S(k)\sim k^{-2}$ for short wave numbers. The condition $S(k)\sim k^{-2}$ is also satisfied for $T<T_c$ if $\rho\sigma=\left[2\gamma_r(\gamma\Delta/\sigma-1)\right]^{-1}$, in  agreement with Eq.\ \eqref{S(0)}. This defines in the $\rho$-$T$ plane a spinodal line or locus of points of infinite isothermal compressibility (within the compressibility route). The spinodal line cannot be extended to temperatures larger than the critical value $T_c$ because  $\left[2\gamma_r(\gamma\Delta/\sigma-1)\right]^{-1}\geq\rho_{\max} \sigma$ if $T\geq T_c$, where $\rho_{\max}$ is the maximum density beyond which the HPA is unphysical at a given temperature (see Appendix \ref{app:appa}).   We further note that the spinodal line only has a lower density (or vapor-like) branch, thus hampering the interpretation of $(T_c,\rho_c)$ as a conventional critical point.

The above features are already suggestive of considering the HPA spinodal line as an artifact of the theory when used in a region
of  parameter space where the  approximation is invalid. Additional support to this view stems from the fact that the HPA keeps predicting a spinodal
line and a critical point even in the SW case ($\epsilon_r/\epsilon_a\to\infty$, $\gamma_r\to 1$), a clearly
incorrect feature. As we shall discuss further below, specialized numerical simulations coupled with  a recent analytical study,\cite{Fantoni10}
strongly support the absence of any phase transition in the present one-dimensional PSW model.

\begin{figure}
\includegraphics[width=\columnwidth]{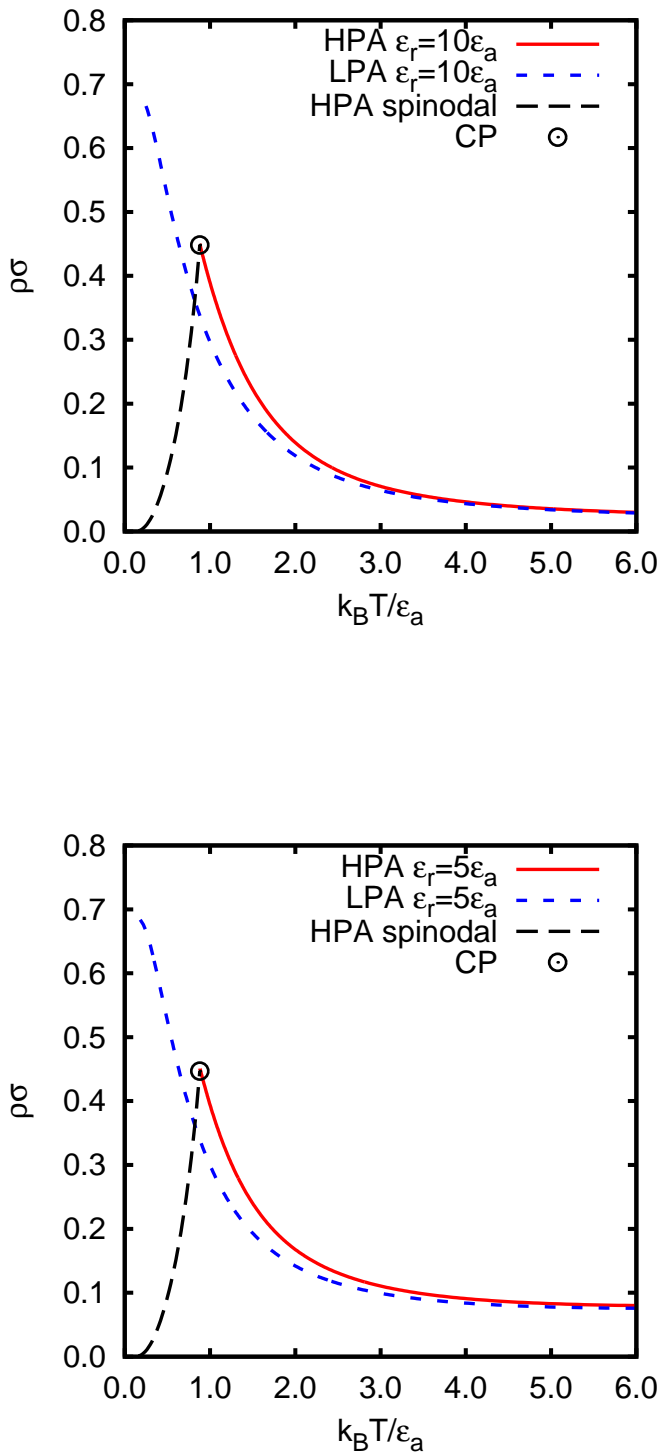}
\caption{Plot of the FW transition line in the $\rho \sigma$ vs
$k_BT/\epsilon_a$ plane with $\Delta/\sigma=1$ and $\epsilon_r/\epsilon_a=10$
 (top panel) and  $\epsilon_r/\epsilon_a=5$
(bottom panel). The long dashed curves are the spinodal lines predicted by the HPA. The FW and spinodal liness meet at the critical point (denoted by a circle).
\label{fig:fig7}}
\end{figure}

In Fig.\ \ref{fig:fig7} we report the comparison between the FW lines, as predicted by the LPA and the HPA, in the  $\rho \sigma$  vs $k_B T/\epsilon_a$ plane for
$\Delta/\sigma=1$ and two different energy ratios $\epsilon_r/\epsilon_a=10$ and $\epsilon_r/\epsilon_a=5$. The spinodal line predicted by the HPA is also included. As said above, the FW and spinodal lines  meet at the critical point. While at high temperatures (above $k_BT/\epsilon_a=3$) there is a remarkable agreement between the two approximations,
deviations occur at lower temperatures.

\begin{figure}
\includegraphics[width=\columnwidth]{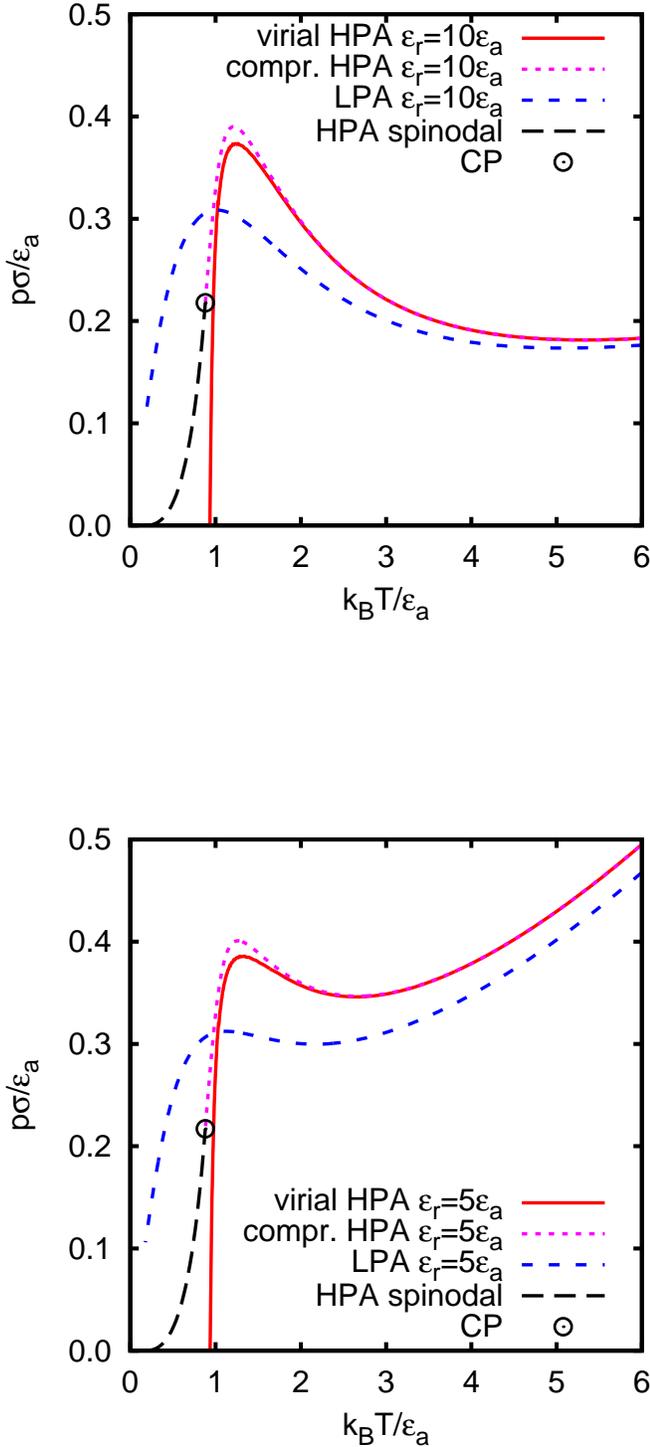}
\caption{Same as in Fig.\ \protect\ref{fig:fig7}, but in the $p\sigma/\epsilon_a$ vs $k_BT/\epsilon_a$ plane. Note that while in the LPA the three routes to the pressure are not
distinguishable one from the other on the graph scale, for the HPA
the difference between the virial and the compressibility route is
noticeable at low temperatures.
\label{fig:fig8}}
\end{figure}

As in the original work by Fisher and Widom,\cite{Fisher69} we also report the FW line in the $p \sigma/\epsilon_a$ vs $k_B T/\epsilon_a$ plane
(see Fig.\ \ref{fig:fig8}), where again we compare the lines as derived from the HPA and LPA schemes. For the energy ratio $\epsilon_r/\epsilon_a=10$
(top panel) we see that the HPA and the LPA give qualitative similar forms of the FW line with a significant deviation at low temperatures, where again the HPA FW line is interrupted at $T=T_c$.
Note that, while all three standard routes give practically identical results within the LPA, the virial(-energy) route in the HPA differs from the
compressibility one at low temperatures (more than $5\%$ for $k_B T/\epsilon_a \lesssim 1.2$). For consistency, the HPA spinodal line is obtained via the compressibility route only.
Similar features occur for the second lower value of the energy ratio, namely $\epsilon_r/\epsilon_a=5$ (bottom panel). Again, the LPA and HPA lines are qualitatively similar, the three routes in the LPA provide indistinguishable results, and the virial and energy routes in the HPA deviate more than $5\%$ for $k_B T/\epsilon_a \lesssim 1.2$. The main distinctive feature in this case $\epsilon_r/\epsilon_a=5$ is  a marked
upswing of the tail of the FW line, absent the previous case $\epsilon_r/\epsilon_a=10$.  This means that, on increasing penetrability
---that is, on decreasing $\epsilon_r/\epsilon_a$--- the transition from oscillatory (above the line) to monotonic (below the line) behaviors occurs at a higher pressure and a higher density for a fixed temperature $k_BT/\epsilon_a$.

Despite the important differences in the steps followed to derive the LPA and the HPA, it is noteworthy that they agree in the qualitative shape of the FW lines (even though the HPA predicts a spurious spinodal line).  It is then reasonable to expect that the true
FW line should interpolate the LPA line at low temperatures with the HPA line at high temperatures.

Finally, we now tackle the issue of the existence of a phase transition for the PSW one-dimensional model. In view of the HPA results
on the seemingly existence of a spinodal line (and hence of a critical point), we here consider the fluid-fluid transition.
As we shall see, our numerics is compatible with the absence of such a transition, thus supporting the view
that the above findings of a spinodal line is indeed a consequence of the application of the HPA to a regime where the theory is not valid.

As the FW line always anticipates the critical point, as remarked,  we can then look for the existence of a fluid-fluid coexistence line in the region predicted by the interpolation
of the LPA and HPA Fisher--Widom lines.
We have carried out extensive simulations of the PSW fluid using Gibbs ensemble Monte Carlo techniques and
employing all standard improvements suggested in the literature.\cite{Panagiotopoulos87,Panagiotopoulos88,Smit89a,Smit89b}
In order to validate our code, we tested it against the case of the one-dimensional SW potential, where exact analytical predictions
for all thermodynamic quantities are available.

We have used up to $1000$ particles and carefully scanned the temperature range  $0.1< k_BT/\epsilon_a<2.0$ and the density range
$0.1 < \rho \sigma < 6$,
as suggested by the FW line (see Fig.\ \ref{fig:fig7}). We have also considered
different values of $\epsilon_r/\epsilon_a$ and $\Delta/\sigma$ for cases giving a significant overlapping probability.
In all the cases we have not found any signature of a fluid-fluid phase separation.

Although the \textit{absence} of a critical transition is always much more difficult to assess as opposed to its presence,
the first scenario is consistent with more than one indication. The first indication stems from a lattice model counterpart of the  one-dimensional PSW
model. This is discussed in Appendix \ref{app:appb}, where the lattice version of the PSW model is constructed following standard manipulations
with the result that no phase transitions are present for \textit{finite occupancy}.
An additional evidence supporting the absence of any fluid-fluid or freezing transition stems from the very recent exact analytical work alluded to earlier,\cite{Fantoni10} that, using the methodology presented in Ref.\ \onlinecite{Kastner08}, concludes that \textit{no phase transitions}
are present for the PSW and PS models in one dimension.

\begin{figure}
\includegraphics[width=\columnwidth]{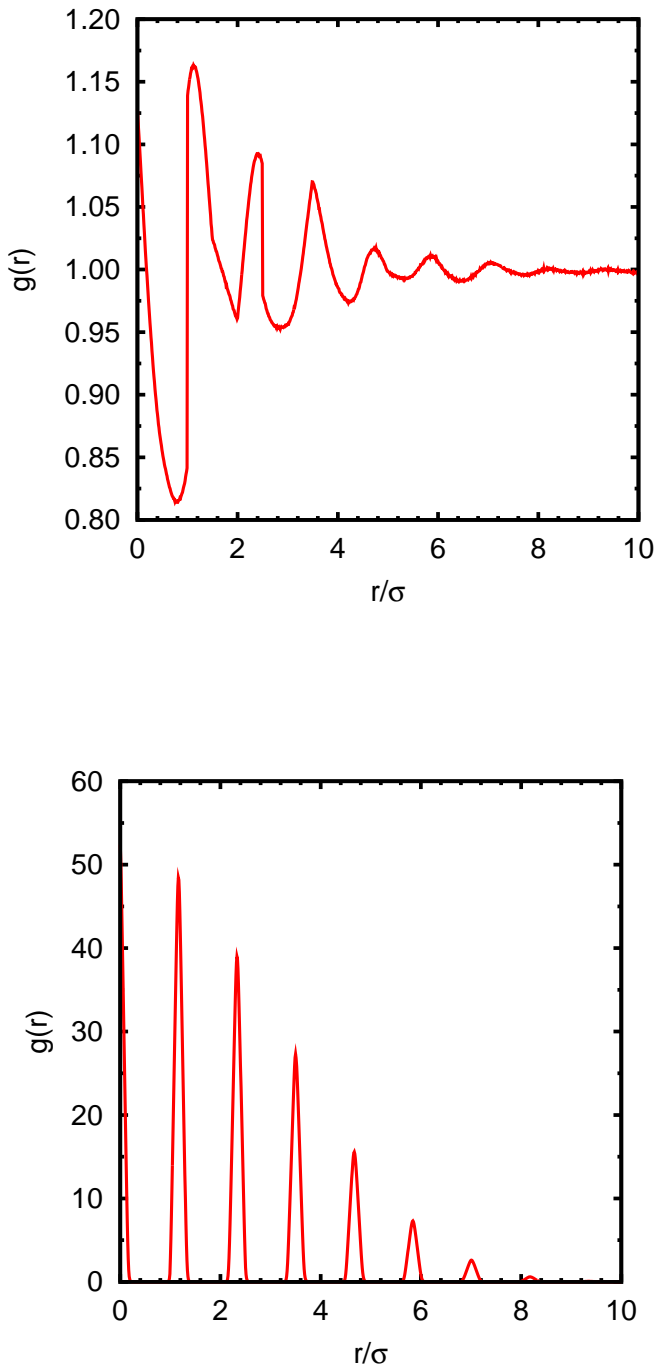}
\caption{Plot of the RDF $g(r)$ obtained from MC simulations for $\epsilon_r/\epsilon_a=2$, $\Delta/\sigma=1.5$, $k_BT/\epsilon_a=10$, and $\rho\sigma=6$ (top panel) and $\rho\sigma=8$ (bottom panel).
\label{fig:blob}}
\end{figure}

In our simulations we have also investigated values of $\epsilon_r/\epsilon_a$ and $\Delta/\sigma$ which violate Ruelle's stability criterion (see Appendix \ref{app:ruelle}) and thus the one-dimensional PSW model is not necessarily stable in the thermodynamic limit.  Here the phenomenology turns out to be much more interesting. For sufficiently low temperatures and sufficiently high densities we observe the formation of a ``blob'' of many-particle clusters (each made of a large number of overlapping particles) having a well defined and regular distribution on the axis, and occupying only a portion of the system length. In the ``blob'' phases the energy per particle grows with the number of particles, thus revealing the absence of a thermodynamic limit. The transition from a ``normal'' phase to a ``blob'' phase if $\epsilon_r/\epsilon_a<2(\ell+1)$, where $\ell$ is the integer part of $\Delta/\sigma$, is illustrated in Fig.\ \ref{fig:blob}. This figure shows the RDF at $\rho\sigma=6$ (top panel) and $\rho\sigma=8$ (bottom panel) for $\epsilon_r/\epsilon_a=2$, $\Delta/\sigma=1.5$, and $k_BT/\epsilon_a=10$. At the lower density the structure of the PSW fluid is qualitatively not much different from that expected if $\epsilon_r/\epsilon_a>2(\ell+1)$ (compare, for instance, with Fig.\ \ref{fig:fig5}). However, the structure at the higher density is reminiscent of that of a solid, except that the distribution of particles does not span the whole length  (here $L =N/\rho= 62.5 \sigma$).
Instead, the particles distribute into a few clusters  regularly spaced with distance $1.17 \sigma$, so that the particles of a given cluster interact attractively with all the particles of the nearest- and next-nearest-neighbor clusters. Note that the number of particles
within any given cluster is not necessarily identical. It is also worth noticing that, in spite of the huge difference in the vertical scales of both panels in Fig.\ \ref{fig:blob}, they are consistent with the
condition $\int_0^{L/2} dr\,h(r)=-1/2 \rho$.

The decay of the peaks of $g(r)$ is mainly due to the lack of translational invariance, i.e., the first and last clusters
have  only one nearest-neighbor cluster, the first, second, next-to-last, and last clusters have only one next-nearest-neighbor cluster, and so on.

We stress that the above phenomenon is specific of bounded potentials, such as  PSW, and has no counterpart in
the hard-core domain. It is then plausible to expect their appearance even in the corresponding three-dimensional versions
of these models where freezing transition (and phase separation for SW) are present but could both be hampered by the presence of
this clustering phenomenon in the region of parameter space where Ruelle's stability criterion is violated.
This would extend the interesting phenomenology already established for the PS case \cite{Likos01b}.  Work along these lines on the three-dimensional case is underway and will be reported elsewhere.

%%%%%%%%%%%%%%%%%%%%%%%%%%%%%%%%%%%%%%%%%%%%%%%%%%%%%%%%%%%%%%%%%%%%%%%%%%%%%%%
\section{Conclusions}
\label{sec:conclusions}
%%%%%%%%%%%%%%%%%%%%%%%%%%%%%%%%%%%%%%%%%%%%%%%%%%%%%%%%%%%%%%%%%%%%%%%%%%%%%%%
In this paper we have completed the study initiated in previous work\cite{Santos08,Fantoni09}
on the penetrable-square-well model in one-dimension. This is a model combining
the three main ingredients present in many colloidal-polymer solutions, namely repulsions, attractions,
and penetrability. While the first two are ubiquitous even in simple fluids, the latter is
a peculiarity of complex fluids where there exist many examples of colloid-polymer systems which
are penetrable (with some energy cost) to some extent, and they involve both steric repulsions
and short-range attractions. This model then captures all these crucial features at the simplest
level of description within an implicit solvent description.

The main new point of this paper was to present an additional and complementary approximation,
denoted as the HPA, valid in regimes complementary to those valid for the low-penetrability scheme
LPA discussed in Ref.\ \onlinecite{Fantoni09}. While the idea behind the LPA was to modify the \textit{exact} relations valid for the one-dimensional
SW fluid ---and in fact for any fluid with a hard core and short-range attraction---
to allow penetrability within some reasonable approximation, the driving force behind the HPA
is the fact that for bounded potential the Mayer function can be made arbitrarily small
by considering sufficiently high temperatures. As a consequence, only the linear chain diagrams need to be retained at each order in the cluster expansion. As it turns out,\cite{Santos09} this is tantamount to considering the celebrated Debye--H\"{u}ckel theory for charged fluids, and we have here considered the
soundness of this approximations at various regimes as compared to specialized MC simulations.
The latter were also compared with other approximations which parallel the entire hierarchy of
approximations in the framework of charged fluids, ranging from the most sophisticated non-linear Debye--H\"{u}ckel
to the simplest mean-spherical approximations. We have assessed the regime of reliability of these approximations
both for thermodynamic and correlation functions by comparison with MC simulations and by internal consistency between
different routes to thermodynamics.

Next we have also discussed the location of the FW line,
separating oscillatory from monotonic behavior in the correlation function, within the HPA and compared with
that obtained from the LPA introduced in previous work.\cite{Fantoni09}
In agreement with previous findings, we find that penetrability enhances the region where correlation functions
have a monotonic regime. The FW derived from the HPA and LPA schemes are found to be in qualitative agreement thus making it
possible the drawing of a line interpolating the high- and low-temperature regimes.

As a final point, we investigated the possibility of a fluid-fluid transition. This possibility arises because
the boundness of the potential renders the van Hove theorem on the absence of phase transition
for one-dimensional model with short-range interactions non-applicable. In fact, the HPA is seen to predict a critical point where the FW line meets a spinodal line. However, this prediction takes place in a region of densities and temperatures where the HPA is not reliable.
A careful investigation using both
NVT and Gibbs ensemble Monte Carlo techniques akin to those exploited in the investigations of the analog problem
for the three-dimensional SW model yields negative results.
These findings are also supported by analytical arguments based on the lattice gas counterpart where the absence
of transition can be motivated by the absence of an infinite occupancy of each site, as well as by an exact analytical proof\cite{Fantoni10} of a no-go theorem
proving the absence of \textit{any} phase transition in this model, which is in agreement with, and beautifully complements, our work.

In our quest for a possible thermodynamic transition in the one-dimensional PSW model we have explored values of the energy ratio $\epsilon_r/\epsilon_a$ and well width $\Delta/\sigma$ for which the stability of the system in the thermodynamic limit is not guaranteed by Ruelle's criterion.\cite{Fisher66,Ruelle69} We have found that, as the temperature decreases and/or the density increases, a transition from a normal fluid phase to a peculiar solid-like phase takes place. The latter phase is characterized by the formation of clusters of overlapping particles occupying a small fraction of the available space and with non-extensive properties. This clustering
transition preempts both the fluid-fluid and fluid-solid transitions.

In view of the results presented here, it would be very interesting to discuss the phase diagram of
the corresponding three-dimensional PSW model.
The phase diagram of the SW model ($\epsilon_r/\epsilon_a\to\infty$) is indeed well established and includes both a fluid-solid transition ---present even in the HS counterpart--- and
a fluid-fluid transition line. The latter is present for any value of $\epsilon_a$ and $\Delta/\sigma$
but is stable against freezing only for  $\Delta/\sigma>0.25$, the depth of the well being irrelevant. \cite{Kranendonk88}
The results presented here strongly suggest the importance of the additional parameter $\epsilon_r/\epsilon_a$.
A first interesting issue would be the Ruelle instability in three-dimensions. A straightforward extension
of the arguments presented in Appendix \ref{app:ruelle}  predicts a guaranteed stability for  $\epsilon_r/\epsilon_a > 12$ (if $\Delta/\sigma<1$),
but the actual onset of the instability cannot be assessed through these arguments. One could expect that for sufficiently high penetrability  (i.e., $\epsilon_r/\epsilon_a<12$ and high density) a phenomenon akin to the ``clustering'' transition found here could be present. In the case $\epsilon_r/\epsilon_a > 12$, where the clustering transition is not expected, the interesting point is to assess the influence of the ratio $\epsilon_r/\epsilon_a$ on the location of the fluid-fluid critical point and coexistence line. All of this opens the possibility of a rich and interesting
phase diagram which would complement that already present in a general class of bounded potentials with no attractive tails.
\cite{Likos01b,Likos07} We note that the high-penetrability
regime is indeed the realm of the HPA presented here, which can be obviously extended to three-dimensions.

Work on the three-dimensional PSW model including the above points and other aspects is underway and will be reported elsewhere.

%%%%%%%%%%%%%%%%%%%%%%%%%%%%%%%%%%%%%%%%%%%%%%%%%%%%%%%%%%%%%%%%%%%%%%%%%%%%%%%
\begin{acknowledgments}
 One of us (RF) wishes to thank M. Kastner for useful discussions and ongoing collaboration. He also gratefully acknowledges the support of the NITheP of South Africa. The support of a PRIN-COFIN 2007B58EAB grant is acknowledged.
The research of A.S. was supported by the Spanish government through grant No.\ FIS2007-60977, partially financed by FEDER funds. AM would like to acknowledge the financial support of the MSMT of the Czech
Republic under Project No. LC512 and the GAAS of the Czech Republic (Grant
No. IAA400720710).
\end{acknowledgments}
%%%%%%%%%%%%%%%%%%%%%%%%%%%%%%%%%%%%%%%%%%%%%%%%%%%%%%%%%%%%%%%%%%%%%%%%%%%%%%%

%%%%%%%%%%%%%%%%%%%%%%%%%%%%%%%%%%%%%%%%%%%%%%%%%%%%%%%%%%%%%%%%%%%%%%%%%%%%%%%
\appendix

\section{Ruelle's stability criterion}
\label{app:ruelle}
%%%%%%%%%%%%%%%%%%%%%%%%%%%%%%%%%%%%%%%%%%%%%%%%%%%%%%%%%%%%%%%%%%%%%%%%%%%%%%%

Let us consider the 1D PSW model characterized by $\epsilon_r/\epsilon_a$ and $\Delta/\sigma$. Let us call $\ell$ the integer part of $\Delta/\sigma$, i.e., $\ell\leq \Delta/\sigma<\ell+1$.
According to Ruelle's criterion, a sufficient condition of
thermodynamic stability is \cite{Fisher66,Ruelle69}
\begin{equation}
U_N(x_1,\ldots,x_N)=\sum_{i=1}^{N-1}\sum_{j=i+1}^N
\phi(|x_i-x_j|)\geq -NB
\end{equation}
for all configurations $\{x_i\}$, where $B$ is a fixed bound.

Given the number of particles $N$, we want to obtain the configuration with the minimum potential energy $U_N$.
Without loss of generality we can see any given configuration  as a set of $M$ clusters $(1\le M\le N)$,  each
cluster $i$ being made of $s_i$   \emph{overlapping} particles
(i.e., any pair of particles of a given cluster are separated a
distance smaller than $\sigma$). In Ref.\ \onlinecite{Santos08} we proved that for a fixed value of $M$ the minimum energy corresponds to $s_i=s=N/M$, all the particles of each cluster being located at the same point and the centers of two adjacent clusters being separated a distance $\sigma$. Therefore, we can restrict ourselves to this class of ordered configurations and use $s$ as the variational variable.

The repulsive contribution to the potential energy is
\begin{equation}
U_N^r(s)=M\frac{s(s-1)}{2}\epsilon_r.
\label{app1}
\end{equation}
To compute the attractive contribution we need to take into account that all the particles of a given cluster interact attractively with the particles of the $\ell+1$ nearest clusters. The total number of pairs of interacting clusters are $(\ell+1)[M-(\ell+1)]+\ell+(\ell-1)+(\ell-2)+\cdots +1= (\ell+1)\left(M-1-\ell/2\right)$. Therefore,
\begin{equation}
U_N^a(s)=-(\ell+1)\left(M-\frac{\ell+2}{2}\right)s^2\epsilon_a.
\label{app2}
\end{equation}
The total potential energy $U_N=U_N^r+U_N^a$ is
\begin{equation}
U_N(s)=N\frac{s-1}{2}\epsilon_r-(\ell+1)\left(N-\frac{\ell+2}{2}s\right)s\epsilon_a.
\label{app3}
\end{equation}
We then see that the value that minimizes $U_N(s)$  is
\begin{equation}
s_*=\frac{N}{\ell+2}\left[1-\frac{\epsilon_r}{2(\ell+1)\epsilon_a}\right].
\label{app4}
\end{equation}
This value is only meaningful if ${\epsilon_r}/{2(\ell+1)\epsilon_a}<1$. Otherwise, $s_*=1$. In summary, the absolute minimum value of $U_N$ is
\begin{widetext}
\begin{equation}
U_N^*=U_N(s_*)=\begin{cases}
 -\frac{N}{2}\left\{\epsilon_r+N\frac{\ell+1}{\ell+2}\epsilon_a\left[1-\frac{\epsilon_r}{2(\ell+1)\epsilon_a}\right]^2\right\}, & {\epsilon_r}<{2(\ell+1)\epsilon_a},\\
-(\ell+1)\left(N-\frac{\ell+2}{2}\right)\epsilon_a,& \epsilon_r>2(\ell+1)\epsilon_a.
\end{cases}
\label{app5}
\end{equation}
\end{widetext}

Therefore, if ${\epsilon_r}>{2(\ell+1)\epsilon_a}$ the potential energy is  bounded from below by  $-NB$ with $B=(\ell+1)\epsilon_a$ and thus the system is stable in the thermodynamic limit. On the other hand, if ${\epsilon_r}<{2(\ell+1)\epsilon_a}$ there exist  configurations that violates Ruelle's
criterion and so the thermodynamic stability of the system is not guaranteed.

%%%%%%%%%%%%%%%%%%%%%%%%%%%%%%%%%%%%%%%%%%%%%%%%%%%%%%%%%%%%%%%%%%%%%%%%%%%%%%
\section{Density expansion of $w(r)$ within the HPA}
\label{app:appa}
%%%%%%%%%%%%%%%%%%%%%%%%%%%%%%%%%%%%%%%%%%%%%%%%%%%%%%%%%%%%%%%%%%%%%%%%%%%%%%
Starting from Eq.\ (\ref{hpa:eq4}) and  for $\rho \gamma_r |\widetilde{f}_{\SW}(k)|<1$, the Fourier transform $\widetilde{w}(k)$
can be expanded in power series as
\begin{eqnarray} \label{appa:eq1}
\widetilde{w}\left(k\right)&=&\sum_{n=2}^\infty \left(\rho \gamma_r\right)^{n-1}\widetilde{f}_{\SW}^n\left(k\right)~.
\end{eqnarray}
Upon inverse Fourier transform one then has
\begin{eqnarray} \label{appa:eq2}
w\left(r\right)&=&\sum_{n=2}^\infty \left(\rho \gamma_r\right)^{n-1}w_n\left(r\right)~,
\end{eqnarray}
where
\begin{eqnarray} \label{appa:eq3}
w_n\left(r\right)&=&\int_{-\infty}^\infty\frac{dk}{2\pi}e^{ikr}\widetilde{f}_{\SW}^n\left(k\right)~.
\end{eqnarray}

Equation (\ref{hpa:eq1}) can be rewritten as
\begin{equation}
\widetilde{f}_\SW(k)=\frac{i}{k}\left[(1+\gamma)\left(e^{ik}-e^{-ik}\right)-\gamma \left(e^{ik\lambda}-e^{-ik\lambda}\right)\right].
\label{9}
\end{equation}
Therefore,
\begin{eqnarray} 
\widetilde{f}_{\text{SW}}^n\left(k\right)&=&\frac{i^n}{k^n}
\sum_{m=0}^n\sum_{p=0}^m\sum_{q=0}^{n-m}
\binom{n}{m}\binom{m}{p}\binom{n-m}{q}\nonumber\\
&&\times(-1)^{m+p+q}
\left(1+\gamma\right)^m
\gamma^{n-m}\nonumber\\ \label{appa:eq4}
&&\times{e^{ik[2p-m+(2q-n+m)\lambda]}}~.
\end{eqnarray}

The origin ($k=0$) is a regular point of $\widetilde{f}_{\SW}(k)$ and hence
of $\widetilde{f}_{\SW}^n(k)$ [but not of each separate term in
Eq.\ (\ref{appa:eq4})], so we can choose to save the point $k=0$ in
Eq.\ (\ref{appa:eq3}) either from above or from below. Here we do it from
above with the result
\begin{equation} 
\label{appa:eq5}
\lim_{\epsilon\to 0^+}i^n\int_{\cal L}\frac{dk}{2\pi}\frac{e^{ikr}}{(k+i\epsilon)^n}=
\frac{(-r)^{n-1}}{\left(n-1\right)!}\Theta\left(-r\right),\quad n>0,
\end{equation}
where the path ${\cal L}$ in the complex $k$ plane goes from $k=-\infty$
to $k=+\infty$ and closes itself on the upper plane if $r>0$ and in
the lower one if $r<0$.
In Eq. (\ref{appa:eq3}) we then find
\begin{widetext}
\begin{eqnarray} \label{appa:eq6}
w_n(r)&=&
\sum_{m=0}^n\sum_{p=0}^m\sum_{q=0}^{n-m}
\binom{n}{m}\binom{m}{p}\binom{n-m}{q}(-1)^{m+p+q}
\left(1+\gamma\right)^m
\gamma^{n-m}\nonumber\\
&&\times\frac{\left[-r-2p+m-\left(2q-n+m\right)\lambda\right]^{n-1}}{\left(n-1\right)!}
\Theta\left[-r-2p+m-\left(2q-n+m\right)\lambda\right]~.
\end{eqnarray}
\end{widetext}
It is interesting to note that $w_n(r)=0$ if $r>\lambda n$. Thus
Eq.\ (\ref{appa:eq2}) can be rewritten as
\begin{eqnarray} \label{appa:eq7}
w(r)&=&\sum_{n=\max\{2,\left[r/\lambda\right]\}}^\infty\left(\rho\gamma_r\right)^{n-1}w_n\left(r\right)~,
\end{eqnarray}
where $[r/\lambda]$ is the integer part of $r/\lambda$.

\begin{figure}
\includegraphics[width=\columnwidth]{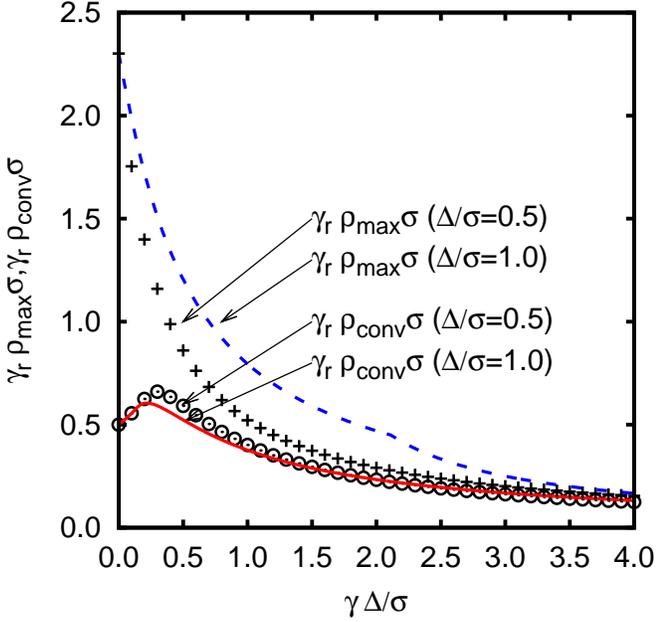}
\caption{Plot of the radius of convergence $\gamma_r\rho_\text{conv}\sigma$  and of the maximum density $\gamma_r\rho_\text{max}\sigma$  versus $\gamma\Delta/\sigma$ for $\Delta/\sigma=0.5$  and $\Delta/\sigma=1$.
\label{fig:fig0}}
\end{figure}

The radius of convergence of the series (\ref{appa:eq7}) depends on temperature and can be
obtained by the same arguments as in the PS case.\cite{Malijevsky06} From
the denominator of Eq.\ (\ref{hpa:eq4}), it follows that the series
converges provided that $\rho<\rho_\text{conv}$, where
\begin{eqnarray} \label{hpa:eq8}
(\gamma_r\rho_\text{conv})^{-1}&=&{\left|\widetilde{f}_{\text{SW}}\right|_{\max}}~.
\end{eqnarray}
Here $\left|\widetilde{f}_{\text{SW}}\right|_{\max}$ denotes the absolute maximum value of $\left|\widetilde{f}_{\text{SW}}(k)\right|$. { From Eq.\ (\ref{hpa:eq1}) if} $\gamma<(\lambda^3-1)^{-1}$ that maximum  corresponds to $k=0$, i.e.,
$\left|\widetilde{f}_{\text{SW}}\right|_{\max}=-\widetilde{f}_{\text{SW}}(0)$,  and so $\gamma_r\rho_\text{conv}=(1-\gamma\Delta)^{-1}/2$. On the other hand, if $\gamma>(\lambda^3-1)^{-1}$ the maximum value  $\left|\widetilde{f}_{\text{SW}}\right|_{\max}$ takes place at $k\neq 0$ and so $\gamma_r\rho_\text{conv}$ must be obtained numerically. For sufficiently large values of $\gamma\Delta$ one has $\left|\widetilde{f}_{\text{SW}}\right|_{\max}=\widetilde{f}_{\text{SW}}(0)$, so that $\gamma_r\rho_\text{conv}=(\gamma\Delta-1)^{-1}/2$ and this coincides with the maximum physical density (see below). Figure \ref{fig:fig0} shows $\gamma_r\rho_\text{conv}$ as a function of $\gamma\Delta$ for two values of $\Delta/\sigma$. In the PS limit ($\gamma\Delta\to 0$) one has $\gamma_r\rho_\text{conv}=\frac{1}{2}$. As the strength of the attractive part of the potential (measured by the product $\gamma\Delta$) increases, the radius of convergence first grows, reaches a maximum, and then decays.

Except when $\gamma\Delta$ is so large that $\left|\widetilde{f}_{\text{SW}}\right|_{\max}=\widetilde{f}_{\text{SW}}(0)$, the maximum $\left|\widetilde{f}_{\text{SW}}\right|_{\max}$ corresponds to a negative value of $\widetilde{f}_{\text{SW}}(k)$ and so the singularity responsible for the radius of convergence is located on the negative real axis. Therefore, $w(r)$ is still well defined beyond the radius of convergence, i.e., for $\rho>\rho_\text{conv}$. On the other hand, analogously to the PS case,\cite{AS04,GK77}  the HPA for the PSW fluid becomes unphysical, at a given temperature,  for densities larger than a certain  value $\rho_\text{max}$ given by the condition
\begin{eqnarray} \label{hpa:eq9}
(\gamma_r\rho_\text{max})^{-1}&=&\widetilde{f}_{\text{SW}}^{\max}~,
\end{eqnarray}
where $\widetilde{f}_{\text{SW}}^{\max}$ is the absolute maximum value of $\widetilde{f}_{\text{SW}}(k)$. Since $\widetilde{f}_{\text{SW}}^{\max}\leq \left|\widetilde{f}_{\text{SW}}\right|_{\max}$, it is obvious that $\rho_{\max} \geq \rho_\text{conv}$.  For sufficiently large values of $\gamma\Delta$ (actually, for temperatures below the critical value $T_c$ defined in Section \ref{sec:FW}) one has $\widetilde{f}_{\text{SW}}^{\max}=\widetilde{f}_{\text{SW}}(0)$, and so  $\gamma_r\rho_\text{max}=(\gamma\Delta-1)^{-1}/2$. In that case, the line of maximum density becomes a spinodal line, as discussed in Section \ref{sec:FW}. Figure \ref{fig:fig0} also includes a plot of $\gamma_r\rho_\text{max}$ as a function of $\gamma\Delta$ for the same two values of $\Delta/\sigma$. Note the kink  of the curve $\gamma_r\rho_\text{max}$ for $\Delta/\sigma=1$  at the critical point ($\gamma_c\Delta/\sigma\simeq 2.11$, $\gamma_r\rho_c\simeq 0.45$), so that  $\gamma_r\rho_\text{max}=(\gamma\Delta-1)^{-1}/2$ (spinodal line) if $\gamma\Delta/\sigma> 2.11$.

%%%%%%%%%%%%%%%%%%%%%%%%%%%%%%%%%%%%%%%%%%%%%%%%%%%%%%%%%%%%%%%%%%%%%%%%%%
\section{The PSW lattice gas}
\label{app:appb}
%%%%%%%%%%%%%%%%%%%%%%%%%%%%%%%%%%%%%%%%%%%%%%%%%%%%%%%%%%%%%%%%%%%%%%%%%%
Consider the PSW model in one dimension. The grand partition function is
\begin{eqnarray}
\label{appb:eq1}
\Xi\left(\mu,L,T\right) &=& \sum_{N=0}^{\infty} \frac{1}{N!} \left(\frac{e^{\beta \mu}}{\Lambda_T} \right)^N Z_N\left(L,T\right),
\end{eqnarray}
where $\Lambda_T$ is the thermal de Broglie's wavelength and
\begin{equation}
Z_N\left(L,T\right)= \int_{L}dr_1\ldots dr_N \exp\left[-\beta \frac{1}{2} \sum_{i \ne j=1}^N \phi\left(\mathbf{r}_i
-\mathbf{r}_j\right) \right].
\label{appb:eq2}
\end{equation}
We now discretize the length $L$ as a sum of $N_c \gg 1$ cells of size $a=L/N_c$ with occupancy $n_{\alpha}$, $\alpha=1,\ldots, N_c$.
The value of $a$ is chosen in the interval $\sigma<a<\sigma+\Delta$, with $\Delta<\sigma$, so that two particles in the same cell are assumed to interact repulsively and two particles in adjacent cells are assumed to interact attractively.
The integral can then approximated as $\int_L d r \approx a N_c$ and the configurational partition function $Z_N$ as
\begin{eqnarray}
Z_N\left(L,T\right) &\approx& a^{N} \sum_{\{n_{\alpha}\}} \delta_{N,\sum_{\alpha} n_{\alpha}} \exp\left[
-\beta \frac{1}{2} \sum_{\alpha \beta =1}^{N_c} \chi_{\alpha ,\beta} \tilde{\phi}_{\alpha \beta} \right]\nonumber\\
&&\times N !.
\label{appb:eq3}
\end{eqnarray}
In Eq.\ (\ref{appb:eq3}) the $\delta$ function accounts for the constraint $N=\sum_{\alpha} n_{\alpha}$ and
the factor $N!$ of the indistinguishability of the particles. Also we have introduced $\chi_{\alpha \beta}$
accounting for multi-interactions among cells and $\tilde{\phi}_{\alpha \beta}$ which is equal to $\epsilon_r$
if $\alpha=\beta$ and equal to $-\epsilon_a$ for nearest-neighboring (nn) cells. Noting that each particle in an $\alpha$-cell
can either interact with $n_{\alpha}-1$ other particles within the same cell or with $n_{\beta}$ particles within
a nn $\beta$-cell, we see that
\begin{equation}
\frac{1}{2} \sum_{\alpha, \beta=1}^{N_c} \chi_{\alpha \beta} \tilde{\phi}_{\alpha \beta} =
-\epsilon_a \sum_{\langle \alpha,\beta \rangle} n_{\alpha} n_{\beta} + \frac{1}{2}\epsilon_r \sum_{\alpha} n_{\alpha}
\left( n_{\alpha}-1\right).
\label{appb:eq4}
\end{equation}
Substituting into Eq.\ (\ref{appb:eq1}) we then find
\begin{widetext}
\begin{eqnarray}
\Xi\left(\mu,L,T\right) &=& \sum_{N=0}^{\infty} \left(\frac{a}{\Lambda_T} e^{\beta \mu} \right)^N
\sum_{\{ n_{\alpha} \}} \delta_{N,\sum_{\alpha}n_{\alpha}} \exp\left[-\beta \frac{1}{2} \sum_{\alpha, \beta} \chi_{\alpha \beta}
\tilde{\phi}_{\alpha \beta} \right].
\label{appb:eq5}
\end{eqnarray}
Because of the $\delta$, the two sums can be inverted and the sum over $N$ can be explicitly carried out thus obtaining
\begin{eqnarray}
\Xi\left(\mu,L,T\right) &=& \sum_{\{ n_{\alpha} \}} \exp\left[-\beta
\left( -\epsilon_a \sum_{\langle \alpha \beta \rangle} n_{\alpha} n_{\beta} + \frac{1}{2}\epsilon_r\sum_{\alpha}  n_{\alpha}
\left(n_{\alpha}-1\right) -\tilde{\mu} \sum_{\alpha} n_{\alpha} \right) \right]
\label{appb:eq6}
\end{eqnarray}
with $\tilde{\mu} = \mu + (1/\beta) \ln(a/\Lambda_T)$.

Assume a finite length $L$ (and hence a finite number of cells $N_c$) with periodic boundary conditions.
The above partition function can then be solved by standard transfer matrix techniques
\begin{eqnarray}
\Xi\left(\mu,L,T\right) &=&  \sum_{\{ n_1,\ldots n_{N_c} \}} \prod_{\alpha=1}^{N_c} A_{n_{\alpha} n_{\alpha+1}} = \operatorname{Tr}A^{N_c},
\label{appb:eq7}
\end{eqnarray}
where we have introduced the matrix
\begin{eqnarray}
\label{appb:eq8}
 A_{n_{\alpha} n_{\beta}} &=& \exp\left[-\beta
\left( -\epsilon_a  n_{\alpha} n_{\beta} +
\frac{\epsilon_r}{4}
\left(n_{\alpha}\left(n_{\alpha}-1\right)+n_{\beta}\left(n_{\beta}-1\right) \right)-
\frac{\widetilde{\mu}}{2}\left(n_{\alpha}+n_{\beta}\right)
\right) \right].
\end{eqnarray}
\end{widetext}
If $N_c$ is \textit{finite}, one then has in the thermodynamic limit
\begin{eqnarray}
\lim_{L \to \infty} \frac{1}{L} \log \Xi\left(\mu,L,T\right) &=& \frac{1}{a}\log \lambda_0,
\label{appb:eq9}
\end{eqnarray}
where $\lambda_0$ is the largest eigenvalue of the matrix. Clearly this is analytic and no phase transitions
are possible for \textit{finite occupancy} $N_c$.

%%%%%%%%%%%%%%%%%%%%%%%%%%%%%%%%%%%%%%%%%%%%%%%%%%%%%%%%%%%%%%%%%%%%%%%%%%%%%%%
% References
%%%%%%%%%%%%%%%%%%%%%%%%%%%%%%%%%%%%%%%%%%%%%%%%%%%%%%%%%%%%%%%%%%%%%%%%%%%%%%%
\bibliographystyle{apsrev}
%\bibliography{1dpsw}

\begin{thebibliography}{99}
%1%
\bibitem{Marquest89} C.~Marquest and T. A.~Witten, J. Phys. (Paris) \textbf{50}, 1267 (1989).
%2%
\bibitem{Likos01} C. N. Likos, Phys. Rep. \textbf{348}, 267 (2001).
%3%
\bibitem{Stillinger76} F. H. Stillinger, J. Chem. Phys. \textbf{65}, 3968 (1976).
%4%
\bibitem{Likos98} C. N. Likos, M. Watzalwek, and H. L\"owen, Phys. Rev. E
\textbf{58}, 3135 (1998).
%5%
\bibitem{Santos08} A. Santos, R. Fantoni, and A. Giacometti,
Phys. Rev. E \textbf{77}, 051206 (2008).
%6%
\bibitem{Fantoni09} R. Fantoni, A. Giacometti, Al. Malijevsk\'y, and
A. Santos, J. Chem. Phys. \textbf{131}, 124106 (2009).
%7%
\bibitem{Malijevsky06} Al. Malijevsk\'y and A. Santos, J. Chem. Phys.
\textbf{124}, 074508 (2006).
%8%
\bibitem{MYS07}
Al. Malijevsk\'y, S. B. Yuste, and A. Santos, Phys. Rev. E \textbf{76}, 021504 (2007).
%9%
\bibitem{Hansen86} J.-P. Hansen and I. R. McDonald \textit{Theory of Simple Liquids}
(Academic Press, Amsterdam, 2006).
%10%
\bibitem{Santos09} A. Santos, R. Fantoni, and A. Giacometti, J. Chem. Phys. \textbf{131}, 181105 (2009).
%11%
\bibitem{VanHove50} L. Van Hove, Physica (Amsterdam) \textbf{16}, 137 (1950).
%12%
\bibitem{Cuesta04} J. A. Cuesta and A. Sanchez, J. Stat. Phys. \textbf{115}, 869 (2004).
%13%
\bibitem{Panagiotopoulos87} A. Z. Panagiotopoulos,   Mol. Phys. \textbf{61}, 813 (1987).
%14%
\bibitem{Panagiotopoulos88} A. Z. Panagiotopoulos, N. Quirke, M.
  Stapleton, and D. J. Tildesley, Mol. Phys. \textbf{63}, 527 (1988).
%15%
\bibitem{Smit89a} B. Smit, Ph. De Smedt, and D. Frenkel,
  Mol. Phys. \textbf{68}, 931 (1989).
%16%
\bibitem{Smit89b} B. Smit and D. Frenkel, Mol. Phys. \textbf{68}, 951 (1989).
%17%
\bibitem{Fantoni10} R. Fantoni and M. Kastner, unpublished (2010).
%18%
\bibitem{Kastner08} M. Kastner, Rev. Mod. Phys. \textbf{80}, 167 (2008).
%19%
\bibitem{Fisher69} M. E. Fisher and B. Widom, J. Chem. Phys. \textbf{50},
3756 (1969).
%20%
\bibitem{AS04}
L. Acedo and A. Santos,   Phys. Lett. A \textbf{323}, 427 (2004).
%21%
\bibitem{Santos07} A. Santos,  J. Chem. Phys. \textbf{126}, 116101 (2007).
%22%
\bibitem{Likos01b} C.~N.~Likos, A.~Lang, M.~Watzlawek and H. L\"{o}wen, Phys. Rev. E
\textbf{63}, 031206 (2001)
%23%
\bibitem{Likos07} C.~N.~Likos, B.~M.~Mladek, D.~Gottwald and G.~Kahl, J. Chem. Phys. \textbf{126}, 224502 (2007).
%24%
\bibitem{note1}
MC results were obtained from a standard NVT ensemble calculation with periodic boundary
conditions. We used $N=500$ particles and $10^6$ MC steps, which were checked to be sufficient for a good accuracy.
%25%
\bibitem{Vega95} C. Vega, L. F. Rull, and S. Lago, Phys. Rev. E \textbf{51}, 3146 (1995).
%26%
\bibitem{Fisher66} M.E. Fisher and D. Ruelle, J. Math. Phys. \textbf{7},
260 (1966).
%26%
\bibitem{Ruelle69} D. Ruelle, \textit{Statistical Mechanics: Rigorous Results}
(Benjamin, London, 1969).
%27%
\bibitem{Kranendonk88} W.~G.~T.~Kranendonk and D. Frenkel, Mol. Phys. \textbf{64}, 403 (1988).
%28%
\bibitem{GK77}
N. Grewe and W. Klein, J. Math. Phys. \textbf{18},  1729, 1735 (1977);
W. Klein and N. Grewe, J. Chem. Phys. \textbf{72},  5456 (1980).
\end{thebibliography}

\end{document}